\newcommand\FINAL[1]{#1}
\newcommand\PLUSCOURT[1]{}

\newcommand\coFINAL[1]{#1}
\FINAL{
  \renewcommand\coFINAL[1]{}
}

\newcommand\coPLUSCOURT[1]{#1}
\PLUSCOURT{
\renewcommand\coPLUSCOURT[1]{}
}

\documentclass[graybox,envcountsame,envcountsect]{svmult}

\newcommand\OK[1]{
}

\usepackage{todonotes}

\newcommand{\olivierOK}[2][]{\OK{\todo[inline,color=red!40,caption={2do}, #1]{\begin{minipage}{\textwidth-4pt}
Olivier:			#2\end{minipage}}} }

\newcommand{\feedbackOK}[2]{\OK{\todo[inline,color=blue!20!white,caption={2do}]{\begin{minipage}{\textwidth-4pt}
#1:			#2\end{minipage}}}}

\newcommand{\feedbackNIOK}[2]{\OK{\todo[inline,color=blue!5!white,caption={2do}]{\begin{minipage}{\textwidth-4pt}
#1:			#2\end{minipage}}}}

\newcommand{\feedbackCOOK}[2]{\OK{\todo[inline,color=blue!1!white,caption={2do}]{\begin{minipage}{\textwidth-4pt}
#1:			#2\end{minipage}}}}

\usepackage{mathptmx}       
\usepackage{helvet}         
\usepackage{courier}        
\usepackage{type1cm}        
\usepackage{makeidx}         
\usepackage{graphicx}        
\usepackage{multicol}        
\usepackage[bottom]{footmisc}

\makeindex             

\usepackage{xcolor}
\usepackage{url}
\usepackage{amsmath,amssymb}
\usepackage{amsfonts}

\usepackage{tikz}
\usepgflibrary{arrows}
\usetikzlibrary{fit}
\usetikzlibrary{decorations.pathreplacing}
\usetikzlibrary{decorations.markings}
\usetikzlibrary{patterns}
\usepackage[utf8]{inputenc}
\usepackage[T1]{fontenc}
\usepackage{subcaption}

\usepackage{hyperref}

\newcommand\R{\mathbb{R}}
\newcommand\Rp{\mathbb{R}_{\geqslant 0}}
\newcommand\N{\mathbb{N}}
\newcommand\Q{\mathbb{Q}}

\newcommand{\myop}[1]{\operatorname{#1}}
\newcommand{\poly}{\operatorname{poly}}
\newcommand{\glen}[1]{\myop{len}_{#1}}
\newcommand{\inorm}[2]{\left\lVert{#1}\right\rVert_{#2}}
\newcommand{\twonorm}[1]{\inorm{#1}{2}}
\newcommand{\infnorm}[1]{\inorm{#1}{}}

\definecolor{darkgreen}{rgb}{0.1,0.6,0.1}
\colorlet{myyellow}{yellow!80!blue}

\newcommand\NEW[1]{\textcolor{blue}{{#1}}}

\newcommand\NEWurl[1]{\NEW{\url{#1}}}

\newcommand\QUID[1]{\coFINAL{\marginpar{auteur: #1}}}

\newcommand\ABOUT[1]{\coFINAL{  {\tiny **} \begin{marginpar}  
    {
      #1} 
\end{marginpar}  }}
\newcommand\TODO[1] { \ABOUT{\textbf{TODO: #1}}}
\newcommand\AFAIRERELIRE[1] { \coFINAL{\begin{marginpar}{\textcolor{blue}{\textbf{A FAIRE
        RELIRE/COMMENTER PAR: #1}}} \end{marginpar}}}
\newcommand\AFAIRERELIREE[1] {
  \coFINAL{\begin{marginpar}{$<${{Envoyé à: #1}}$>$} \end{marginpar}}}

\newcommand\TEXTEVOLE[1]{  \coFINAL{begin{quote} #1 \end{quote}  } }

\begin{document}

\title*{A Survey on Analog Models of Computation 
}
\titlerunning{A Survey on Analog Models of Computation. 
}
\author{Olivier Bournez and Amaury Pouly}
\institute{Olivier Bournez \at Olivier Bournez, Campus de L'Ecole
  Polytechnique, 91128 Palaiseau Cedex France, \email{bournez@lix.polytechnique.fr}
\and Amaury Pouly \at MPI-SWS, E1 5, Campus, 66123 Saarbr\"{u}cken, Germany \email{pamaury@mpi-sws.org}}
%
%

\maketitle


\newcommand\TEXTE{
\feedbackOK{Jérémie}{"in analogy": est-ce que c'est correct/ce que l'on veut dire ? L'expression "by analogy with"
exists et est synonyme de "by comparing with". Il me semble que "by analogy" serait plus correct.
\textcolor{red}{BY}
}
We present a survey on analog models of computations. Analog
can be understood both as computing by analogy, or
as working on the continuum. We consider both approaches, often
intertwined,  with a point of view mostly oriented  by
computation theory. 
}

\abstract*{
\TEXTE
}
\abstract{
\TEXTE
}

\feedbackOK{Emmanuel Jeandel}{ Salut,
(d) Je ne suis pas complètement d'accord avec la division
continu/discret.

La question principale est de savoir si $\{0,1\}^NN$ est considéré
continu ou discret, et plus généralement si on considère un espace
zero-dimensionel comme continu ou discret.

De ce que je lis dans l'intro "by opposition to spaces like $N^k$ ([...]
or the set of configurations of a model such as a Turing machine)",
vous considerez que $\{0,1\}^NN$ est discret.

Je ne suis pas forcément d'accord, mais dans le contexte de l'article,
c'est une distinction raisonnable.
Ca met donc les automates cellulaires, même sur des entrées infinies,
dans la rubrique "discrète", c'est un choix que vous assumez
pleinement.

Mais le raisonnement dans 3.1 en prend un coup:

"objects involved in discrete computations (...) have a finite
representation".

Ceci n'est plus vrai, puisque la configuration d'un automate
cellulaire (ou d'une machine de Turing partant d'une entrée
quelconque) n'est pas descriptible de façon finie, et tous les
arguments que vous développez dans ce même paragraphe s'appliquent
tout aussi bien à ces modèles.
\textcolor{red}{OLIVIER}
}

\feedbackOK{Erik Winfree}{
Your chapter looks like a great contribution to the community — but also it’s quite a mouthful, so I didn’t have time to look at it closely this morning.  I will share with you some minor thoughts upon looking at figure 1, nonetheless.  I would not be surprised at all if you already take care of these points in the body of the text, just without them being reflected in the figure.  So please don’t take this as criticism.  Just a thought.

I was surprised to see the relative emptiness of the “discrete space / continuous time” quadrant.  There are many models that have an underlying semantics expressed in the language of continuous-time markov chains (CTMCs) because it is so natural to derive from physics.  For example, stochastic chemical reaction networks (Gillespie / CME) would fall here, in my view.   So would most flavors of asynchronous cellular automata.  Likewise, many models of molecular self-assembly.

Perhaps part of the awkwardness is that people often mental group
together variants of a model that operate under different semantics.
For example, I am aware of Hopfield’s associative memory neural
networks with discrete-time/discrete-state,
continuous-time/continuous-state, and continuous-time/discrete-state
formulations of the dynamics.   Similarly, standard cellular automata
and asynchronous cellular automata, CRNs can come with stochastic
discrete (CTMC) semantics or with deterministic continuous (ODE)
semantics — and another major flavor for CRNs changes discrete SPACE
to continuous SPACE, i.e. reaction-diffusion models that use PDEs.
One could even view PDEs themselves as being a continuous SPACE
variant of cellular automata, or visa versa.
}

\feedbackOK{Erik Winfree}{
As concluding notes about fig 1, I would argue that spiking neural networks are a particularly interesting class of continuous-time/continuous-state model.   Also, although I would argue that “chemical reaction network” is misplaced in the discrete-time/discrete-state quadrant, it would be natural to place Petri Nets there (although, they too come in many flavors!).   Finally, another aspect that might be worth clarifying this this figure (you could use color!) is that some models have discrete specifications (e.g. cellular automata), while others allow continuous parameters in their specification (e.g. real-valued weights in neural networks, real-valued rate constants in CRNs, etc).  There are obviously some major issues related to allowing real-valued specification of a system.

Which, presumably, you do touch upon in the parts of the paper that I did not read.

Cheers,
Erik

\textcolor{red}{OLIVIER}
}

\feedbackNIOK{Erik Winfree}
{

I am glad that you found my feedback of some use.  

While most of the theoretical work on spiking neural networks (especially that of Wolfgang Maass) was from a decade ago or more, there appears to be a resurgence of practical implementations.  For example

Merolla, Paul A., et al. "A million spiking-neuron integrated circuit with a scalable communication network and interface." Science 345.6197 (2014): 668-673.
Diehl, Peter U., and Matthew Cook. "Unsupervised learning of digit recognition using spike-timing-dependent plasticity." Frontiers in computational neuroscience 9 (2015).

Best regards,
Erik
\textcolor{red}{OLIVIER}
}

\feedbackOK{Ulf Hashagen}{

It seems to me that it would be very useful to incorporate some of the
standard historical reference works on the history of analog computers
and some literature on special topics as well.  Please forgive me that
I will give this in a form of a list of very short commentaries and
references. I have not published the material about the history of
analog computing in Germany (I have spoken in Lille about), but parrts
are included in a general overview which is at the end of this list. 

1. General literature on history of analog computers: 

\cite{care2010technology} a) Care, C.: Technology for Modelling: Electrical Analogies, Engineering Practice and the Development of Analogue Computing. Springer, 2010  [BEST BOOK! ]

\cite{mickelson1963analog} = b)  Bromley, A. G.: Analog Computing Devices. In: Aspray, W. (Ed.): Computing Before Computers., Iowa State University Press, 1990, 156-199

\cite{williams1997history}=c) Williams, M. R.:  A History of Computing Technology. IEEE Computer Society Press, 1997 [CHAPTERS ON ANALOG COMPUTING]

2. History of analog instruments

\cite{fischer2002instrumente}, \cite{fischer1995instrumente}=
a) Fischer, J.: Instrumente zur Mechanischen Integration. Ein Zwischenbericht. In: Schütt, H.-W. and Weiss, B. (Eds.): Brückenschläge. 25 Jahre Lehrstuhl für Geschichtfe der exakten Wissenschaften und der Technik an der Technischen Universität Berlin 1969-1994., Engel, 1995, 111-156
b) Fischer, J.:  Instrumente zur Mechanischen Integration II. Ein (weiterer) Zwischenbericht. In: Schürmann, A. and Weiss, B. (Eds.)Chemie -- Kultur -- Geschichte: Festschrift für Hans-Werner Schütt anlässlich seines 65. Geburtstages., GNT-Verlag, 2002, 143-155

3. History of Differential Analyzer

\cite{owens1986vannevar}= a) Owens, L.: Vannevar Bush and the Differential Analyzer: The Text and Context of an Early Computer. Technology and Culture, 1986, 27, 63-95 [STANDARD ARTICLE! ]

4. ENIAC

\cite{haigh2016eniac}=
a) Haigh, T.; Priestley, M. and Rope, C.: ENIAC in Action. MIT Press,
2016 = 

5. History of (Analog) Computing in Germany by Ulf Hashagen 

\cite{hashagen2011rechner}=
a) Hashagen, U.:  Rechner für die Wissenschaft: "`Scientific Computing"' und Informatik im deutschen Wissenschaftssystem 1870--1970. IN: Hashagen, U. and Hellige, H. D. (Eds.): 
Rechnende Maschinen im Wandel: Mathematik, Technik, Gesellschaft. Festschrift für Hartmut Petzold zum 65. Geburtstag., Deutsches Museum, 2011, 111-152

Moreover I should add that some of your historical  judgements (about ENIAC, Pascaline etc.) are conflicting with the literature in the history of computing. Please don't hesitate to contact me again, if you any questions. 

Best regards, Ulf Hashagen

\textcolor{red}{OLIVIER À VÉRIFIER}
}

\feedbackOK{John Tucker}{
\textcolor{red}{AMAURY: AJOUTER UN PEU}

Dear Olivier and Amaury,

Thank you for the draft of this excellent survey! Thank you for asking for comments, too.

I have some new suggestions, clarifications and replacement/extra references that you could make use of.  They appear below.

I can send some further attachments of some of the papers.

I can send further comments on the text if you wish.

Best wishes,

John

++++++++++++++++++++++++++++++++++++++++++++++++++

Introduction

First, relevant to your Introduction and your Figure 1 is this paper: perhaps you could look at it ... I attach a copy for convenience.

It proposes a theory of synchronous concurrent algorithms (SCAs) as to unify modelling and reasoning about many spatially distributed parallel algorithms --- neural nets, cellular automata, systolic algorithms, coupled map lattices, numerical grids, etc. It has continuous data in focus.  

\textcolor{red}{AMAURY: OK}
\cite{thompson2009unifying} = B C Thompson, J V Tucker and J I Zucker, Unifying computers and dynamical systems using the theory of synchronous concurrent algorithms, Applied Mathematics and Computation, 215 (2009) 1386-1403.

Section 2: Some Analog Machines

\textcolor{red}{AMAURY: PEUT ETRE LE PREMIER}
I noticed that you wrote about the Blum-Shub-Smale model in 2.5.

I wondered if you would mention that finite computability models have been generalised to abstract algebras long before the the BSS model by Harvey Friedman, John Shepherdson, Erwin Engeler and myself. I considered applications to topological spaces in:

J V Tucker, Computing in algebraic systems, in F R Drake and S S Wainer (eds.), Recursion Theory, its Generalisations and Applications,  London Mathematical Society Lecture Note Series 45, Cambridge University Press, Cambridge, 1980, pp. 215-235.

In particular, subsequently, Jeff Zucker and I re-developed  all of basic computability theory for many sorted algebras, with a particular attention to computation on topological algebras (and, naturally, the reals).

There is a major introduction to the general theory, with a detailed history, is:
\textcolor{red}{AMAURY: PEUT ETRE, ET DIRE QUE ETENDU AU CONTINU}
J V Tucker and J I Zucker, Computable functions and semicomputable sets on many sorted algebras, in S Abramsky, D Gabbay and T Maibaum (eds.), Handbook of Logic in Computer Science. Volume V: Logical and Algebraic Methods, Oxford University Press, 2000, pp. 317-523.

and we have a recent update with plenty of new material on continuous data: 

J V Tucker and J I Zucker, Generalizing computability to abstract algebra, Turing's ideas - their significance and impact.  A volume commemorating the centennial of Alan Turing's birth, (editors) G. Sommaruga and T. Strahm, Birkhauser, 2016, pp. 127-160.

Computation on continuous data using abstract  programming models (free of any representation of the structures) is special as it demands a beautiful use of non-determinism for search and choice: 

J V Tucker and J I Zucker, Abstract versus concrete models of computation on partial metric algebras, ACM Transactions on Computational Logic, 5 (4) (2004) 611-668.
    
Concrete computability models (based on representations of the structure) are quite different.  Stoltenberg-Hansen and wrote a survey:

\cite{stoltenberg2008computability} V Stoltenberg-Hansen and J V Tucker,  Computability on topological spaces via domain representations, in S B Cooper, B Löwe and A Sorbi (eds), New Computational Paradigms: Changing Conceptions of What is Computable, Springer-Verlag, 2008, pp.153-194.

Section 5. Theory of Analog Systems

5.1.2

Thank you for including  Jeff Zucker's and my work on networks of analog units.  Your citation of [289] concerns continuity and has a sequel on computability that could be added in the []:

\cite{TZ14} J V Tucker and J I Zucker, Computability of operators on continuous and discrete time streams, Computability, 3 (2014) 9-44. 
 Doi: 10.3233/COM-14024 

You might note that in the papers we apply the general results in case studies, by actually working though the semantic modelling of mechanical systems (taken from an old Analog Computing textbook).

Section 6. Analysing the Power and Limits of Analog Systems

6.2

Thank you for mentioning the Beggs-Costa-Tucker programme on physical oracles.

Here are some suggestions.

In addition to [41], you could cite that we recently published a survey of the main theory as it has settled down to date:

\cite{ambaram2017analogue} Tânia Ambaram, Edwin Beggs, José Félix Costa, Diogo Poças, J V Tucker, An Analogue-digital Model of Computation: Turing Machines with Physical Oracles, in: Advances in Unconventional Computing, Book series: Emergence, Complexity and Computation, Vol. 22, Springer-Verlag, 2016, 73-115.

I attach a copy for convenience.

The key thing about [41] is the statement of a "Church-Turing Thesis' that includes analog data taken from physical measurement and that strictly and dramatically extends the Turing Barrier.

An important paper that is missing is this one that brings together the many two-sided -- and so main types -- of examples of physical oracles in a full axiomatisation:

\cite{beggs2012axiomatizing} E.J. Beggs, J.F. Costa and J.V. Tucker, Axiomatising physical experiments as oracles to algorithms, Philosophical Transactions of the Royal Society A 28 (2012) 370, 3359-3384. doi:10.1098/rsta.2011.0427

It could be cited on the top of page 30, for example.

For interest, our programme on physical oracles began over a decade ago with:

\cite{beggs2007experimental}E.J Beggs, J.V Tucker, Experimental computation of real numbers by Newtonian machines, Proceedings of the Royal Society Series A, 463 (2007) 1541-1561

\cite{beggs2008computational}  E J Beggs, J F Costa, B Loff, and J V Tucker, Computational complexity with experiments as oracles, Proceedings of the Royal Society Series A, 464 (2008) 2777-2801.

\cite{beggs2009computational} E J Beggs, J F Costa, B Loff, and J V Tucker, Computational complexity
with experiments as oracles. II. Upper bounds, Proceedings of the
Royal Society Series A, 465 (2009), 1453-1465.

}

\section{Introduction} {\QUID{olivier}


\feedbackNIOK{
Liesbeth DE mol}{In relation to the use of the distinction
analog/digital: there is an upcoming paper on this by Ronald Kline in
which he gives historical context to the distinction and shows that
our current distinction was not “straightforward” from the
start. Amongst others, there were the criticisms from Licklider during
a Macy conference where he points out that “analog” in the sense of
“in analogy to” combined with the assumption that analog refers to
continuous (in contrast to the digital) does not work because one can
also have discrete models...
\textcolor{red}{A IGNORER}
}

\feedbackCOOK{Bernd Ulmann}
{
 p. 1: "...and most of the analog machines that were historically built were 
actually hybrid." Althrough hybrid computers, i.e. combinations of analog and 
digital (stored program) computers were conceived quite early, I would guess 
that the majority of analog computer were more or less purely "analog" i.e.
the only "digital" portion with respect of value representation being two-state
was the control unit for controlling the integrator modes.
\textcolor{red}{A IGNORER}
}

\feedbackOK{Bernd Ulmann}
{
 Did you read about the work of Glenn Cowan? He implemented a VLSI analog 
computer back in 2005, and a newer implementation was recently done by 
Ning Guo (\url{https://www.ning-guo.com/publications}), both working for Yannis
Tsividis.
\textcolor{red}{OLIVIER}
}

There is a clear ambiguity about the sense of the word \emph{analog}
when talking about analog computations.  Nowadays, it is often
understood as being the opposite of \emph{digital}: the former is
working on continuous quantities, while the latter is working over
discrete values, typically bits or words. However, historically,
\emph{analog computation} got its name from computation by analogy,
i.e. by systems built in such a way that they evolve exactly as the
system intended to model or simulate \cite{maclennan2009analog,
  LivreAnalogcomputing}. These two understandings are orthogonal and
various machines analog in both or one and not the other sense have
been conceived \cite{LivreAnalogcomputing}. Notice also that even
\emph{discrete} vs \emph{continuous} is not a clear dichotomy, and
most of the analog machines that were historically built were actually
hybrid \cite{LivreAnalogcomputing}.


We will mainly focus on models of computation, with a point of view possibly
oriented by computation theory (computability, complexity, models of
computation).


All considered models can be described as particular
dynamical systems:  a dynamical system is mathematically defined as
the action of a subgroup $\mathcal{T}$ of $\R$ on a space $X$, i.e. by
a function (a \emph{flow}) $\phi: \mathcal{T} \times X \to X$ satisfying the
following two equations: 
\begin{equation}
\phi(0,x)=x \label{flow:one}
\end{equation}
\begin{equation}
\phi(t,\phi(s,x))=\phi(t+s,x). \label{flow:two}
\end{equation}

Function $\phi(t,x)$ is intended to give the position at time $t$ of
the system if started at position $x$ at time $0$, and the above
equations simply
express expected properties.

Subgroups $\mathcal{T}$ of $\R$ are known to be either dense in $\R$
or isomorphic to the integers. In the first case, the time is said to be
\emph{continuous}, in the latter case, \emph{discrete}.

A dynamical system is often alternatively described by some space $X$ and some
function $f: X \to X$. Indeed, in the discrete time case, giving
$\phi$ is equivalent to giving $f: X \to X$ with $f(x)=\phi(1,x)$: the
trajectory starting from some state $x_0$ then corresponds to the
iterations of $f$ on $x_0$ from Equation \eqref{flow:two}. In the
continuous time case, not all dynamical systems correspond to
differential equations, but at soon as $\phi$ is continuously
differentiable, a case covering a very wide class of systems in
practice, we can write $y^\prime=f(y)$ where
$f(y) = \left.  \frac{d}{dt} \phi(t,y) \right|_{t=0}$. Giving $\phi$
is then also equivalent to giving a function $f: X \to X$: the trajectory
starting from some state $x_0$ corresponds to the solution of the
Initial Value Problem  (IVP) $y^\prime=f(y), y(0)=x_0$.

We can then classify models according to their space $X$: we will
mainly focus in this chapter on the case where the space $X$
involves real numbers, i.e. is \emph{continuous}.
Typically $X=\R^n$ or $X=\R^n \times \N^k$ or can be encoded naturally
in similar spaces.  We will say that such a space is \emph{continuous}, by
opposition to spaces like $\N^k$ (or the set of words over a given
alphabet, or the set of configurations of a model such as a Turing
machine) that would correspond to a \emph{discrete} space.  Discrete
time and space corresponds to classical computability and complexity
and are not intended to be covered by this chapter. We will nevertheless
consider some unconventional discrete time and space models, such as
population protocols or chemical reaction networks, as they are
unconventional and turn out to be closely related to analog models as
shown by various recent results.

\begin{figure}
\begin{center} 
\begin{tabular}{|ll||c|c|}
\hline
 & Space & Discrete & Continuous \\
Time &   &          &            \\
\hline
\hline
Discrete 
&&  \PLUSCOURT{\cite{Tur36}}\coPLUSCOURT{Turing}   machines     & \PLUSCOURT{Discrete time \cite{Hopfield84} }
Neural networks (Section \ref{sec:nn1})   \\
&&  \PLUSCOURT{\cite{Church36}} Lambda calculus        &  Deep
                                                         Learning
                                                         models
                                                         (Section
                                                         \ref{sec:nn1})
  \\
&&  \PLUSCOURT{\cite{Kleene36}} Recursive functions                  &
                                                                       \PLUSCOURT{\cite{BSS89}}
                                                                       \coPLUSCOURT{Blum
                                                                       Shub Smale}
                                                                       machines  (Section
                                                         \ref{sec:bss})

\\
&&   \PLUSCOURT{\cite{Post46}} Post systems                &  
Hybrid Systems (Section \ref{sec:hybrid}) 
\\  
&&    Cellular automata                & Natural Computing Influence
                                         Dynamics (Section
                                         \ref{sec:chazelle}) \\
&&Stack automata  & \PLUSCOURT{\cite{Durand-Lose05}} Signal 
machines (Section \ref{sec:durand})
\\
&&  Finite State Automata                &Continuous Automata (Section
                            \ref{sec:continuousautomata}) \\
&& Population Protocols (Section \ref{sec:populationprotocol}) &
                                                                 \vdots  \\ 
&& Chemical Reaction Networks  (Section \ref{sec:SCRN}) &  \\ 
&& Petri Nets & \\ 
&&     \vdots                     &  \\
\hline
Continuous &&
Boolean Difference Equation models \PLUSCOURT{(\cite{BDEI})}                          &
                                                  \PLUSCOURT{\cite{Sha41}}\coPLUSCOURT{Shannon's}
                                                  GPACs and
                                                  (Section \ref{sec:gpac}) \\
            &&                & \PLUSCOURT{Continuous time \cite{Hopfield84}}\coPLUSCOURT{Hopfield's} neural 
networks (Section \ref{sec:nn}) \\
          &&               & Physarum 
                             computing (Section
                             \ref{sec:physarum})  \\

          &&               & Reaction-Diffusion Systems  (Section
                             \ref{sec:reaction-diffusion})  \\

           &&               & Hybrid Systems (Section
                              \ref{sec:hybrid})   \\
           &&               & \PLUSCOURT{\cite{AD91}} Timed automata (Section
                              \ref{sec:timedautomata}) \\

         &&               & Large Population Protocols  (Section
                              \ref{sec:lpp}) \\
         &&               & Black Hole Models   (Section
                              \ref{sec:blackhole}) \\

      &&               & Computational Fields   (Section
                              \ref{sec:computationalfields}) \\

          &&               & \PLUSCOURT{\cite{Moo95b}} $\R$-recursive functions  (Section
                              \ref{sec:rrecursive}) \\
          &&               &  Spiking neuron
                             models (Section
                              \ref{sec:nn1}) \\
           &&               & \vdots  \\
\hline
\end{tabular}
\caption{A tentative classification of some computational models, according to
  their space and time.}
  \label{fig:dyn_sys_classification}
\end{center}
\end{figure}

A classification of some of the models considered in this chapter
according to this discrete/continuous time/space
dichotomies  is provided by Figure
\ref{fig:dyn_sys_classification}. This classification is not perfect
and debatable. For example, cellular automata (that we will consider
later on as a spatial model)  are here considered as
evolving over a discrete space. But there may also be considered as
evolving over 
$\{0,1\}^\N$ (i.e. Baire's space), and we agree that there is no
fundamental difference between $\R$ and $\{0,1\}^\N$. As another example, the discrete-space /
continuous-time quadrant is rather empty, but we agree that many
models have an underlying semantics that can be expressed in the
language of continuous-time Markov chains, and for example stochastic chemical
reaction networks could also fall
in this quadrant. Actually, many models come indeed in various flavours
and could be turned into various quadrants: For example, asynchronous cellular
automata can arguably be considered as being continuous time and discrete
space, or continuous time and continuous space.

Notice that many quantum models of computation can also be considered possibly
as analog and would fit to this description: as we have to make
choices, we decided that they are out of the scope of the current
chapter. Notice that analog models, in particular models in the
continuum, can be seen as fitting in some way in between classical
and quantum models. Several results about quantum models are
deeply using the fact that the underlying space is the field
$\mathbb{C}$ of complex numbers, while analog models can sometimes
emulate some of the constructions by working over
$\mathbb{R}^2$. Similarly, computations with quantum models sometimes
assume measurements to be possible at unit cost, while this is not
considered as possible in most analog models. Discussing analog models
helps to understand in that spirit the aspects which are closely coupled
to quantum world from the others.

Since most of the other chapters of this book are dedicated to
the computable analysis framework, we will intentionally not discuss this approach in the current chapter.

Previous surveys on the field of analog computation include Pekka
Orponen's \cite{Orponen}, Olivier Bournez and Manuel
L. Campagnolo's \cite{CIEChapter2007} focusing on
continuous time computation, and Bruce J.  MacLennan's chapter
on ``Analog Computation'' of the \emph{Encyclopedia of complexity and
  systems science.}  \cite{maclennan2009analog}.
A recent very instructive book about
the history of analog machines has recently been published \cite{LivreAnalogcomputing}. There is also an extensive
literature, mostly forgotten today, about analog machines dating to
the times where most of machine programming was analog.

\AFAIRERELIREE{Bernd Ulmman + MacLennan}

\section{Various Analog Machines and Models}

\subsection{Historical Accounts}{\QUID{Olivier}}\label{sec:historical_analog_machines}

\AFAIRERELIREE{Bernd Ulmman}

As we said, historically, analog computation was mainly referring to
computation by analogy.  Very instructive quotations
supporting this claim can be found in \cite{LivreAnalogcomputing} and
\cite{maclennan2009analog}.


Actually, several of the historical first computers presented as among
the first ever built \emph{digital} computers turn out to be analog
in this above primary sense. This includes the ENIAC, which interestingly
stands for \emph{``Electronic Numerical Integrator and
Computer'}'. The ENIAC is often said to be \emph{``programmed''} but the term
\emph{``wired''} would be closer to reality, since this sytem mainly consisted of
a large collection of arithmetic machines. 

It may also help to understand that the term \emph{computer} was
historically referring to a person who carried out calculations or
computations. From the middle of the 20th century only, the word
started to refer to a \emph{machine} that carries out computations.
Determining which systems can actually be considered as computers is a very
intriguing question, related to deep philosophical questions out of
the scope of the present chapter.

We however list here some of the first ever built machines that can 
be classified as analog: This includes 
Blaise Pascal's 1642 \emph{Pascaline}, as well as 
Johann Martin Hermann's 1814 \emph{Planimeter} (a simple device to compute
surfaces based on Green's theorem),
or
Bill Phillips's
1949 \emph{MONIAC} (\emph{Monetary National Income Analogue Computer}, a 
machine that was using fluidic logic to model the behavior of an
economy). 
The \emph{Antikythera mechanism}, discovered close to the greek island of
Antikythera in 1901, dated from earlier than 87 BC, whose purpose was
to predict astronomical positions and phenomena, is also often
considered as an analog computer.
However, even if these machines were clearly computing various quantities or
data and may or may not be called computers, we admit that even the question
of whether such machines can be classified
as \emph{analog} is debatable. Indeed, many articles or books
consider them as such nowadays (see e.g. \cite{LivreAnalogcomputing}),
but these statements are conflicting with the literature in the history of
computing from historians. 

However, with no contest, the first truly programmable (\emph{``general purpose''}
using Shannon's 1941 terminology) analog computer is Vannevar's Bush
1931 \emph{Differential Analyser}, which is the topic of the next section
due to its historical and fundamental importance.

Further detailed historical accounts about analog machines and computations can be found in
the recent monograph \cite{LivreAnalogcomputing}. Its author, Bernd Ulmann also maintains an
informative web site with instructive pictures and videos
\cite{AnalogComputerMusuem}. \AFAIRERELIREE{Bernd Ulmman + MacLennan}

The history of analog computation and devices, is also discussed by 
historians.  For general literature on history of analog computers,
refer to  
\cite{care2010technology}, 
\cite{mickelson1963analog}
or 
\cite{williams1997history}. For references related to history of analog
instruments, see \cite{fischer2002instrumente},
\cite{fischer1995instrumente}. The history of Differential Analyser  is
discussed in 
\cite{owens1986vannevar}. See also \cite{ulfhashagen} and
\cite{hashagen2011rechner}  for national accounts about the
developments of differential analysers. 
Historical accounts for ENIAC can be found
in 
\cite{haigh2016eniac}. 
Recently published accounts also include monograph
\cite{rojas2002first}, the analysis of the work of Douglas R. Hartree
\cite{durand2016douglas} , Charles Babbage \cite{durand1992charles} or
of particular devices or techniques
\cite{durand2010planimeters,durandmail} before differential
analysers. 

\AFAIRERELIREE{Marie-José Durand-Lose, Ulf Hashagen (et lui demander
  des vrais choses citables)} 

\subsection{Differential Analysers}{\QUID{Amaury} }\label{sec:differential_analyzer}

\AFAIRERELIREE{Bernd Ulmman + Historiens rencontrés à Lille}
\AFAIRERELIRE{Marie-José Durand-Richard}

The probably best known universal continuous time machine
is the \textit{Differential Analyzer (DA)}, built for the first time in
1931 at the MIT under the supervision of Vannevar Bush \cite{Bush31}. The idea of assembling integrator devices to solve differential
equations dates back to Lord Kelvin in 1876 \cite{Tho76}. Kelvin was looking for a
faster way to compute the harmonics of a function using its Fourier transform,
with applications to tidal and meteorological observations. He came up with the
idea of using a rotating disc-cylinder-globe system to compute the integral of a product: this is
essentially the fundamental operation of the Differential Analyzer, although it
took over 50 years to solve the mechanical problems involved in this machine.

The first DAs were entirely mechanical, and later became electronic:
see \cite{Bow96, LivreAnalogcomputing}. Their primary purpose was to solve differential equations,
especially the ones coming from problems in engineering. By the 1960s, differential analysers were
progressively discarded in favor of digital technology. 
Many accounts on the history and applications of these machines can be found in \cite{LivreAnalogcomputing}.

\subsubsection{Mathematical Model of the Differential Analyzer: the GPAC }{\QUID{Amaury}}\label{sec:gpac}

The \emph{General Purpose Analog Computer} (\emph{GPAC}) was introduced by Shannon
in 1941 as a mathematical idealization of the Differential Analyzer \cite{Sha41},
while he was working on this machine at the MIT to get money for his
studies. A GPAC is basically a circuit made up of a finite
number of units, described in Figure~\ref{fig:gpac_circuit}, that are
interconnected by wires, possibly with loops (retroactions) on some of
the wires.  A
function $f: \R \to \R$ is said to
be \emph{generated} if it corresponds to the value read on some of the wires of
the circuit as a function of time.  

The model was later
refined in a series of papers \cite{PER74,LR87,GC03,Gra04}.  Indeed, a GPAC circuit may not
define a unique function (it could have no or several solutions) which
is problematic. An arguably more modern presentation of the GPAC is to
use ordinary differential equations instead of circuits. The
equivalence of GPAC circuits with polynomial ordinary differential equations will be discussed
in 
Section~\ref{sec:theory_generable_functions}. 

\AFAIRERELIREE{Daniel Graça, Felix Costa}
\AFAIRERELIRE{ Jonathan Mills}

Rubel also introduced the \emph{Extended Analog Computer} (\emph{EAC}) as an extension of the GPAC
\cite{Rub93}. The EAC features a broader class of operations such as partial differentiation
and restricted limits, allowing to solve boundary value problems. Rubel's motivation for
introducing the EAC was that the GPAC was ``a limited machine''. This claim has been
somewhat changed since then (see Section~\ref{sec:theory_generable_functions})
and the EAC has seen few theoretical developments. Although Rubel envisioned the EAC
as a ``purely conceptual machine'', there is ongoing research to implement it,
with some currently working prototypes \cite{Mills95, Mills05, Mills2008}.

\begin{figure}[h]
\begin{center}
 \setlength{\unitlength}{1200sp}%
\begin{tikzpicture}
 \begin{scope}[shift={(0,0)},rotate=0]
  \draw (0,0) -- (0.7,0) -- (0.7,0.7) -- (0,0.7) -- (0,0);
  \node at (.35,.35) {$k$};
  \draw (.7,.35) -- (1,.35);
  \node[anchor=west] at (1,.35) {$k$};
  \node at (.35, -.3) {A constant unit};
 \end{scope}
 \begin{scope}[shift={(4,0)},rotate=0]
  \draw (0,0) -- (0.7,0) -- (0.7,0.7) -- (0,0.7) -- (0,0);
  \node at (.35,.35) {$+$};
  \draw (.7,.35) -- (1,.35); \draw (-.3,.175) -- (0,.175); \draw (-.3,.525) -- (0,.525);
  \node[anchor=west] at (1,.35) {$u+v$};
  \node at (.35, -.3) {An adder unit};
  \node[anchor=east] at (-.3,.525) {$u$};
  \node[anchor=east] at (-.3,.175) {$v$};
 \end{scope}
 \begin{scope}[shift={(0,-2)},rotate=0]
  \draw (0,0) -- (0.7,0) -- (0.7,0.7) -- (0,0.7) -- (0,0);
  \node at (.35,.35) {$\times$};
  \draw (.7,.35) -- (1,.35); \draw (-.3,.175) -- (0,.175); \draw (-.3,.525) -- (0,.525);
  \node[anchor=west] at (1,.35) {$uv$};
  \node at (.35, -.3) {A multiplier unit};
  \node[anchor=east] at (-.3,.525) {$u$};
  \node[anchor=east] at (-.3,.175) {$v$};
 \end{scope}
 \begin{scope}[shift={(4,-2)},rotate=0]
  \draw (0,0) -- (0.7,0) -- (0.7,0.7) -- (0,0.7) -- (0,0);
  \node at (.35,.35) {$\int$};
  \draw (.7,.35) -- (1,.35); \draw (-.3,.175) -- (0,.175); \draw (-.3,.525) -- (0,.525);
  \node[anchor=west] at (1,.35) {$w=\int u\thinspace dv$};
  \node at (.35, -.3) {An integrator unit};
  \node[anchor=east] at (-.3,.525) {$u$};
  \node[anchor=east] at (-.3,.175) {$v$};
 \end{scope}
\end{tikzpicture}
\end{center}
\caption{Basic units used in a GPAC circuit.}
\label{fig:gpac_circuit}
\end{figure}
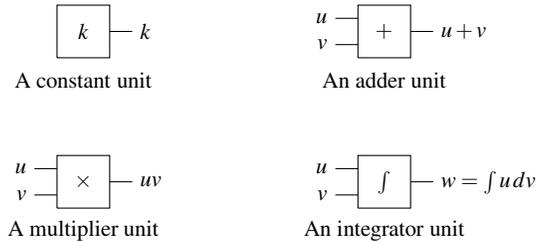

\subsection{Neural Networks and Deep Learning Models }{\QUID{Olivier}}
\label{sec:nn1}

\feedbackOK{Bernd Ulmann}
{
 If we can (I think we can :-) ) agree that at the heart of an analog computer 
is the fact that its operation is determined by its structure and not by an 
analog computer, it might be interesting to have a look at Google's Tensor
Flow which allows one to easily define a dataflow machine which then operates
on n-dimensional datastructures at once. TF might prove quite usable to define
the structure of an analog computer for solving a particular problem. Maybe 
this could be a missing link towards an automatically reconfigurable analog
computer. :-)
\textcolor{red}{OLIVIER DOIT PARLER DE TENSOR FLOW}
}

\AFAIRERELIREE{Jeremie Cabessa + Hava Siegelmann + Orponen si répond}
\AFAIRERELIREE{Wolfgang Maass}

In the 80's and 90's, \emph{artificial neural networks} gave birth to a
renewal of interest in analog computations. This enthusiasm declined until recently and
the success of \emph{deep learning} in several impressive applications in
various fields of artificial intelligence, such as speech
recognition, image recognition and natural language processing. The first
machine able to beat all best professional \emph{Go} players is based
on  deep learning technology.

Current deep learning models may have a
rather complex architecture built from various modules, 
but most of these modules are essentially artificial neural
network models.  
An artificial neural network (ANN) consists of many simple, connected
processors called \emph{neurons}.  The state of each neuron is given
by some real number called its activation value. Designated \emph{input neurons}
get their activation values from the environment. All other neurons
evolve by applying a composition of a certain one-variable function
$\sigma$ (usually a sigmoid) with an affine combination of the
activations of the neurons to which they are connected. Finally,
specific neurons, considered as \emph{output neurons}, may trigger actions on the
environment: see \cite{bishop1995neural} for an overview.

Most of the work related to (deep) artificial neural networks is
nowadays devoted to finding architectures and weights that make an ANN
exhibit a desired behavior in a given context of application
A popular library to describe
corresponding architectures is 
\emph{Tensor Flow} \cite{tensorflow}. This is basically a library to
describe such architectures, and hence can be considered in many
aspects as a library for particular analog computations. 

A very popular and successful method to determine suitable weights
for solving a given problem is  \emph{back-propagation}, which is 
reinforcement learning technique based on a gradient descent method
applied to an error function expected to be minimized over the learning
set, see \cite{bishop1995neural}. 

Applying any gradient descent method requires a differentiable error function with respect to all involved
parameters. The very large number of applications of deep learning
techniques have recently led to the emergence of various other analog models of
computation. All these models have in common to be differentiable
end-to-end versions of models inspired by classical computability. This
includes the popular  \emph{Long Term Short Memory} (\emph{LTSM}) architecture \cite{hochreiter1997long}, or the so-called
\emph{Differentiable Neural Computers} \cite{graves2016hybrid}, or the
\emph{Neural Turing machines} \cite{graves2014neural}. The underlying
principle of these constructions is to extend  an artificial neural
network by coupling it to an external memory resource. This external
resource can also be a stack as in  the so-called \emph{Neural Network Pushdown Automata}
\cite{sun1993neural} and \emph{Neural Stack machines}
\cite{grefenstette2015learning}.

The models discussed previously all work in discrete time, but models with
a continuous time dynamics have also been considered, such as  \emph{symmetric Hopfield networks}.  
The convergence behavior of these networks has been used in various applications such as associative
memory, or combinatorial optimization problems after first
applications in 
\cite{Hopfield84}.

Models based on  \emph{spiking neurons} have been claimed to be
biologically more realistic \cite{MB98}. Various coding methods exist to
interpret the outgoing spike train as a real value number either
relying on the frequency of spikes, or the timing between spikes. This
yields to various ways of encoding information and the computational power
of spiking neuron models has been investigated in  series of papers:
see
\cite{SOSurvey}, \cite{MB98} or \cite{ghosh2009spiking} for
surveys.
Recent years have seen practical implementations \cite{merolla2014million} as well as their use to
solve practical problems efficiently \cite{diehl2015unsupervised}.





\subsection{Models from Verification
}{\QUID{Olivier}}\label{sec:models_for_verification}
\label{sec:hybrid}

\feedbackNIOK{Laurent Doyen}{Pour la décidabilité des automates hybrides, je suggérerais "What's
decidable about hybrid automata?"

Reference = \cite{HKPV95}

 qui contient les preuves, ou le papier
survey "Discrete abstractions of hybrid systems". Ces deux papiers
couvrent ce qu'on aborde sur l'(in)décidabilité dans notre
du chapitre du handbook.

Reference = \cite{alur2000discrete}
\textcolor{red}{OLIVIER A BAFFER}
}

\feedbackNIOK{Goran Frehse}{
Pour les résultats classiques les citations de Laurent sont bien. Je peux ajouter un petit article récent sur les système hybrides à 2 variables, qui trace la frontière de décidabilité :
\url{http://folk.uio.no/gerardo/infcomp2012.pdf}

Reference = \cite{asarin2012low}

Pour des raisons que j'ignore, Springer repousse la date de
publication du Handbook depuis trois ans; a un moment les données sur
leur site étaient plus complets que maintenant. J'ai trouvé une liste
des chapitres ici (tout en bas de la page), mais ce n'est peut-être
pas à jour:
\url{
https://www.bookdepository.com/Handbook-Model-Checking-2017-Jr-Edmund-M-Clarke/9783319105741
}
\textcolor{red}{OLIVIER}
}

\feedbackNIOK{Laurent Fribourg}{
En ce qui concerne la partie reliée à la vérification en particulier, elle me convient personnellement très bien du point de vue de l'idée (a posteriori) que j'aurais pu m'en faire.

Peut-etre  qu'il faudrait là (ou peut-etre avant ?) mentionner les tres interessantes tentatives de donner une semantique formelle à SimuLink de Benoit Caillaux et al. en utilisant
l'arithmetique non-standard
(par ex: Albert Benveniste, Timothy Bourke, Benoît Caillaud, Marc
Pouzet. Non-Standard Semantics of Hybrid Systems Modelers. Journal of
Computer and System Sciences (JCSS), 78(3):877-910, 2012.)
Reference = \cite{benveniste2012non}

ainsi que les travaux de Edward Lee (bon survey dans les slides
"Models of Time for CPS") qui pointent les tres grandes difficulté de modéliser, par exemple, un "simple" pendule de Newton avec des equations differentielles standard.

Reference = \cite{neher2011interval}
\textcolor{red}{OLIVIER}
}

\feedbackNIOK{Laurent Fribourg}{ Oups, j'ai oublié aussi de te dire, concernant la derniere ligne de la Section 3.4, qu'il faudrait peut-etre mentionner les travaux dans
la lignée de Ramon Moore sur l'Interval Arithmetic
(voir par ex.: Interval Methods and Taylor Model Methods for ODEs
Markus Neher, Dept. of Mathematics)
\textcolor{red}{OLIVIER A VOIR}
}

The development of algorithms and techniques for verification or
control of so-called \emph{hybrid systems} or \emph{cyberphysical systems} have also generated several lines
of research related to analog computation. 
 \emph{Hybrid} and  \emph{cyberphysical
systems} have in common to mix discrete evolutions, often a digitally engineered controller,
with continuous dynamics that often comes from the environment or from some natural
continuous variables that the controller acts upon.



\AFAIRERELIREE{Laurent Doyen + Goran Frehse + autres?}
There is an extensive literature on the \emph{hybrid automata} modeling
approach to determine the exact frontier between decidability and
non-decidability for reachability properties, according to the type of
dynamics, guards, and resets allowed.  The
same holds for the frontier about applicability of techniques coming
from control to prove properties about these systems such as
stability. 
 Providing a complete panorama on this literature is out of
the ambition of the current chapter.  Classical survey references are
\cite{HKPV95,alur2000discrete}. See also \cite{CIEChapter2007} for
discussions. 
%

This literature is the source of many  \emph{dynamic undecidability}
results to which we will come back in Section \ref{sec:undec_dyn_sys}. 

Hybrid systems are known to exhibit the so-called \emph{Zeno's
  phenomenon}. In short, they may happen to have an unbounded number of
discrete transitions in a bounded (continuous) time.  In this context, it is classical to
distinguish various types of Zeno's behaviour: \emph{Chattering Zeno}
vs \emph{Genuine Zeno}, following \cite{ames2005characterization}.  The first type can often be eliminated and
corresponds more to an artefact of the model, which can be avoided by
considering an appropriate notion of solution, while the second is
more problematic to detect and harder to avoid in
simulation \cite{ames2006there}. We will come back in Section
\ref{sec:zeno} to Zeno's phenomenon in the wider perspective of space
time contraction phenomena in analog systems. 

\AFAIRERELIREE{Joelle Ouaknine + James Worell}
\AFAIRERELIRE{ Coauteurs?}
Despite some promising early results in the field, open problems even for very simple classes of
systems still remain. In particular, the decidability of the
reachability problem for  \emph{piecewise affine maps on the real line} is a
famous open problem, with some partial progress such as
\cite{Potapov05, bell2016decidability}. Several other questions seem to
be closely related to that issue \cite{asarin2012low}.  The
decidability of  \emph{point to hyperplane}  reachability for
discrete time linear systems, known as the Skolem problem, is famously open.
The \emph{continuous time} version
  of \emph{point to hyperplane} has recently been shown to be related to conjectures from
transcendental number theory \cite{chonev2015skolem}. Hardness
of  \emph{recurrent reachability for continuous linear dynamical
  systems}  of
low dimensions has been investigated in \cite{chonev2016recurrent}.

Several recent results have also focused on the hardness of
\emph{bounded time versions} of reachability problems. The complexity of
the reachability problem has been characterized for hybrid automata in
\cite{brihaye2011reachability}, and for piecewise affine systems in \cite{ben2013mortality} and \cite{JournalRP}.
Complexity of problems or methods from control theory have
also been explicitly derived \cite{ahmadi2013complexity,
  ahmadi2013stability, ahmadi2013switched,ahmadi2014complexity}.

In relation with other chapters of this handbook, we mention that the
recursive analysis approach has also been explored. Computability of reachable and invariant
sets in the framework of computable analysis have been investigated for continuous time systems \cite{Col05}
and  for hybrid systems \cite{Collins05}.

A method to approximate hybrid systems with a polynomial hybrid
automaton, i.e. a Taylor approximation, has also been proposed
\cite{lanotte2007taylor}, as well as interval methods and
Taylor model methods for ODEs \cite{neher2011interval}. There have been attempts
to provide a formal semantics to Simulink based on non-standard analysis tools \cite{benveniste2012non}.
More generally, providing models of systems using ordinary differential equations is
an hard task in practice: See e.g. very instructive discussions in
\cite{CoursALee} about all the difficulties in modeling a simple system such as
Newton's pendulum.


\TODO{Existe: 
 29. Verification of hybrid systems. Handbook of model
 checking. Laurent Doyen, Goran Freshe, George J. Pappas, and André
 Platzer. Encyclopedia/Handbook  of Model Checking. In preparation.}





\subsubsection{Timed Automata }{\QUID{Olivier}}
\label{sec:timedautomata}

\AFAIRERELIREE{Patricia Bouyer} \emph{Timed automata} \cite{AD94}  can be considered as a
restricted version of hybrid systems for which decidability of
reachability holds. They can also be considered as an extension of
finite automata with clocks.  The model has clear practical applications
and is at the heart of several computer tools for verification: see
e.g. recent survey  \cite{BFLMOW-hmc17}. 

From a
more fundamental point of view, timed automata can be seen as language
recognizers \cite{AD94}, and there has been various attempts to
generalize concepts from finite automata theory to this
framework. This includes closure properties of recognized languages
\cite{AD94},   pumping lemmas \cite{Beauquier98} as well as
variants of Kleene's theorem \cite{Asa_,ACM96,AsaCasMal02,BouPet99,BouyerP02,AsarinD02}. For a recent
survey about timed automata, see \cite{BFLMOW-hmc17}.



\subsection{Blum-Shub-Smale's Model }{\QUID{Olivier}}
\label{sec:bss}

\AFAIRERELIREE{Christine Gassner + Klaus Meer}

\feedbackOK{Klaus Meer}{Thanks for citing many of my papers. If you like
  to include the work by others you might think about a real version
  of Toda's theorem by Basa and Zell. 
PAS TENU COMPTE: Or the many results from
  Burgisser and Cucker
}

\feedbackOK{Klaus Meer}{
For [220], I would prefer the journal version in `Information and
computation''. Title on Ladner (majuscule)
\textcolor{red}{OLIVIER}}

The \emph{Blum-Shub-Smale (BSS) model} \cite{BSS89}  has been introduced as a
discrete time model of computation over the reals in order to discuss
hardness of problems in algebraic complexity. In the initial
presentation of the model, operations of the field
$\R$ are assumed to be realizable at unit cost, leading to classes such
as $P_\R$ and $NP_\R$, with complete problems such as the
existence of a real zero of a given polynomial. 
 Later, it has been
generalized to other fields or rings with extended or
restricted operations \cite{BCSS98}, or to an abstract model over
arbitrary logical structures \cite{Poi95}. Notice that classical 
discrete computability models have been generalized to abstract
structures in various ways in parallel, and also before
\cite{BSS89}: See e.g. \cite{friedman1971algorithmic}, \cite{Mos83} or
\cite{TZ00}.

The obtained computability and complexity theory subsumes classical
discrete computation theory since the latter can be seen as the specific
case of logical structures with a finite domain. 

There is an extensive literature on the related computation theory. Many results have
been obtained in this model, mostly studying the corresponding
complexity classes, and their relations with classical questions in
computability theory, or providing lower or upper complexity bounds on
various problems based in this framework: See \cite{BCSS98, Poi95}.

Recent results include non-trivial generalizations to this framework
of Toda's theorem \cite{basu2012complex,basu2010polynomial} or the PCP theorem
\cite{baartse2017algebraic}, interactive proofs
\cite{baartse2016real}, Ladner's result \cite{meer2012ladners},
as well as separation of degrees \cite{gassner2013strong,gassner2010separation}.

The model is different in spirit from most others discussed in
this section, as it is usually not considered to be an
attempt at modeling analog machines in a realistic way, but rather as a
mathematical tool. It has proven to be relevant for the discussion of
lower bounds in algebraic complexity, or for some classical
questions from complexity theory in a wider generalized setting, 
where sometimes complexity classes such as $P$ and $NP$ can be
separated. The model has also clear connexions with the generalized
finite automata models discussed in Section \ref{sec:automatameer}.



\subsection{Natural Computing}
\label{sec:chazelle}

The interest in unconventional models of computation, and in particular for natural computing, has
revived interest in analog computing.

\subsubsection{Dissipative Influence Dynamics}{\QUID{Amaury (premier jet d'olivier)}}

\AFAIRERELIREE{Bernard Chazelle} The framework of \emph{natural algorithms},
has motivated a series of works about models of \emph{influence systems} and
their computational capabilities. A manifesto in favor of the fact that the study of
natural systems can benefit from an algorithmic perspective has been published
in \cite{chazelle2009natural}. This has motivated the
exploration of several models of influence dynamics such as the
Hegselmann-Krause's model. In particular, bounds on the time required
by a group of birds to stabilize in a standard bird flocking model
have been established in \cite{chazelle2014convergence}. Turing
completness and almost sure asymptotic periodicity of diffusive influence
systems have also been obtained in \cite{chazelle2015diffusive}. For a general discussion on the
merits and challenges of an algorithmic approach to natural
algorithms, see \cite{chazelle2015algorithmic}.

\subsubsection{Physarum Computing }
\label{sec:physarum}
\AFAIRERELIREE{Andrew Adamatzky}

{\QUID{Olivier}}
\emph{Physarum polycephalum} is a slime mold that has been shown to be
able to solve various natural problems: this includes
realization of boolean logic gates, implementation of delay in
computing circuits, geometry computations such as Voronoi diagrams or Delaunay triangulations, or
computation of shortest paths \cite{adamatzky2015atlas}. Models of
various aspects of its behaviour have been established. Convergence
proofs and complexity bounds for computing shortest paths have been
investigated \cite{becchetti2013physarum}, based on mathematical model
for the slime's behavior proposed in the form of a coupled system of
differential equations \cite{tero2007mathematical}. Implementation
of Kolmogorov-Uspensky in biological substrate has also been
investigated \cite{adamatzky2007physarum}.  For a survey about these fields of research,
see \cite{adamatzky2015atlas}.










\subsubsection{ DNA and Molecular Computing}{\QUID{Olivier}}



{\QUID{Olivier (initialement prévu: Amaury)}}





Areas such as \emph{systems biology} aim  at understanding
complex biological processes in terms of their basic mechanisms at the
molecular level.  Many attempts of applying concepts and
tools from theoretical computer science to this framework (logic,
algebra, etc..) have been investigated, with numerous
successes and concrete software systems: see \cite{fages2014cells}.

However, even if the primary purpose of these fields was to explain concrete biological features,
various approaches have considered computations by chemical reactions as a programming
tool. The idea is to consider computations by reactions as programs to
solve various tasks, as nature does in the context of cell biology, but not restricting to this context.
In particular, various attempts to relate this concept of
programmation to computation theory have been proposed.  In most of these
approaches, the underlying principle is to get inspiration from concrete chemical phenomenon in
order to derive abstract  models of computation which are potentially usable. This includes the model of
biomolecular computation \cite{hartmann2010programming}, the \emph{Chemical Abstract Machine}
\cite{BB92tcs}, \emph{Membrane Computing} models \cite{PR02tcs}, \emph{Biochemical
Ground Form} process algebra's approaches \cite{CZ10mscs}, DNA based
computation models \cite{QSW11dna}, or \emph{Chemical Reaction Networks}
(CRN) discussed in Section \ref{sec:SCRN}.

Simulations of Turing machines has been demonstrated in several of the
above mentioned articles, both at an abstract level or
concretely. 
A very challenging question is to
compare the actual implementations in nature of some of the tasks to
other possible more efficient ones that could be derived
theoretically \cite{CMSB17}. 

For a presentation of natural computing models and results not covered in this chapter,
see \cite{rozenberg2011handbook,MembraneBook}. In this chapter, we will only briefly discuss
the case of DNA computing (below) as well as that of CRNs (Section \ref{sec:SCRN}).


{\QUID{Olivier}}
DNA computing started with a work \cite{Adl94} which proposed to solve the
directed Hamiltonian path problem on a graph using
DNA as well as enzymes to implement the
computation. Molecular computation had been investigated in various
ways before popular Adlmenan's article (see
e.g. \cite{conrad1990molecular}), including the idea of using DNA
computing to implement computations of formal language theory \cite{Head87}. This has been extended later on to other
known NP-complete problems \cite{Lipton94}, even if the required resources have
been demonstrated to be unrealistic. Approaches based on evolutionary
computation to control resources have also been proposed
\cite{stemmer1995evolution}.  The field developed meanwhile in many
impressive ways. Visible facts include the development of a
programmable molecular computing machine \cite{lovgren2003computer},
the demonstration of the use of DNA as digital storage medium encoding
a 5.27-megabit book \cite{church2012next}, or the construction of the
analog of a DNA transistor \cite{bonnet2013amplifying}.







\subsection{Solving Various Problems Using Dynamical
  Systems}{\QUID{Olivier}}

\feedbackNIOK{Keijo Ruohonen}{
Second, you've really covered in a nicely readable way a very extensive collection of cases of analog computing in various senses. I found these very interesting, there were many quite new to me. If you still have space for another one, I’d suggest the mechanical computer models, the ‘Billiard ball computer’ and others, investigated in the 70s and 80s by such names as

Edward Fredkin
Tommaso Toffoli
Charles Bennett
Norman Margolus

among others. Some references:

Fredkin and Toffoli: Conservative Logic. Int. J. Theor. Phys. 21 (1982), 219–253
Bennett: Logical reversibility of computation. IBM J. Res. and Development 17 (1973), 525–532
Bennett: The Thermodynamics of Computation – A Review. Int. J. Theor. Phys. 21 (1982), 905–940
Margolus: Physics-Like Models of Computation. Physica D 10 (1984), 81–95

This topic was quite popular at the time. Not so much anymore, I guess, but some of the relevant concepts have turned out to be important in quantum computing, too, reversible universal computing (actually discovered much earlier by Yves Lecerf in another context), Toffoli gates etc.

These computing models are physically ideal: no friction, perfect elasticity, perfectly machined parts etc. Especially, gravitational effects are ignored. Newtonian many-body dynamics is of course very complex, and such big names as

Jeff Xia
Richard McGehee
John Mather
Donald Saari

have shown that even without black holes the dynamics still has its Zenoan phenomena, which might make it possible to have super-Turing computation in a classical setting (as has been speculated e.g. by Frank Tipler in his oddball book ‘The Physics of Immortality’). Some references:

Margolus: Physics-Like Models of Computation. Physica D 10 (1984), 81–95
Saari and Xia: Off to Infinity in Finite Time. Notices of the AMS 42 (1995), 538–546
Mather and McGehee: Solutions of the Collinear Four-Body Problem which Become
Unbounded in Finite Time. Lecture Notes in Physics 38 (1975), 573–597

\textcolor{red}{OLIVIER}
}


Several authors have shown that certain, possibly discrete, decision or optimization
problems such as graph connectivity or linear
programming, can be solved by specific continuous dynamical
systems. Some examples and references can be found in the papers
\cite{Smi98,VSD86, 
  Bro88, Ref6-BFFS03,HM94a,HSF02,moore2011nature,blakey2010model}.
This has links with various mechanical computer models (e.g. ``billard
ball computers'') investigated by several papers in the 70s and 80s,
with discussions about physical limits to computations such as
thermodynamic reversibility
\cite{Ben73,fredkin2002conservative,bennett1982thermodynamics,margolus1984physics}. Relevant
concepts have turned out to still be important in quantum computing
(e.g. Tofolli gates, etc.). This is also related to discussions about
various phenomena possibly capable of hypercomputation
\cite{mather1975solutions,saari1995off}: see in particular Section
\ref{sec:blackhole} about Black Hole computations for more recent
references. 

Observe that, if analog computation is to be understood in the sense
of computing by analogy, then almost any historical analog
machine can be considered as falling under this framework. In his monograph,
Ulmann develops with sometimes great and
instructive details some of the machines for particular
problems such as finding zeros of a polynomial, linear algebra,
optimization and simulation \cite{LivreAnalogcomputing}. In more than 70 pages various
historical  applications in about 20 fields such as Mathematics, Physics, Mechanics,
Geology and Economics are described.

%


The question of whether some of the discrete problems could actually be
solved faster using continuous methods is very intriguing: this is the
object of Section \ref{sec:hyper}.

Many dynamical systems of the form $H^\prime=[H,[H,N]]$, where the
notation $ [B, L]$ stands for $BL-LB$, have been shown to be
continuously solving some particular discrete problems
\cite{Bro88,Bro89} such as sorting lists, diagonalizing matrices and
linear programming problems. One key property is that this
equation is equivalent to some gradient flow on the space of
orthogonal matrices. Many examples in this spirit are discussed in details in \cite{helmke2012optimization}.
More recently a system of nonlinear differential equations
of a similar form to sort numbers fed to the input has been
investigated \cite{gladkikh2016study}. 

Notice that several discrete time algorithms or methods have some
analog equivalent. This includes \emph{Newton's method} which
leads to \emph{Newton's flow dynamics} for finding roots. \emph{Gradient descent}
methods (this includes the very popular Backpropagation method for
Neural Networks) can also be considered as the (explicit) Euler's
discretization 
method of a continuous flow, the so-called \emph{gradient's flow}.

The use of analog methods for solving $k$-SAT problems has been
investigated in \cite{ercsey2011optimization}: the problem is mapped
to an ordinary differential equation about which some properties are
established, such as chaotic transience of trajectories above a constraint
density threshold and fractality of the boundaries between the basin of
attraction. The system is stated to always find solutions in polynomial
continuous time, but at the expense of exponential fluctuations in its
energy function \cite{ercsey2011optimization}. An attempt to physically implement the proposed
dynamic has been proposed \cite{yin2016efficient}. Previous statements that $k$-SAT can be formulated
continuously can be found in
\cite{gu1999optimizing,nagamatu1996stability,wah1997trace}. 

Notice that polynomial-size continuous-time Hopfield nets have been
proven able to simulate PSPACE Turing machines \cite{SO2003}: This
implies that even ODEs with Lyapunov-function controlled dynamics can
actually do much more than solving NP-complete problems in some sense.

\subsubsection{Interior Point Methods }{\QUID{Olivier}}

\feedbackOK{Klaus Meer}{Perhaps it is a bit confusing that you mention
  Khachigan in relation with interior point methods. As far as I
  understand, the ellipsoid method is not an interior point method

I also think that in relation with ITPM the work by J. Renegar as to
be cited and so on
\textcolor{red}{OLIVIER A CLARIFIER}
}

A particular class of methods falls very naturally into this framework:
the so-called \emph{interior point methods}, which correspond to a particular
class of algorithms to solve linear and nonlinear convex optimization
problems. The principle, already proposed by John von
Neumann, and very popular in the 1960s, is to build a continuous system
whose trajectories are evolving in the interior of the feasible
region. A common method to guarantee evolution in the feasible regions
was the use of \emph{barrier functions} acting as a potential energy. By
making these functions tend to infinity on the boundary, the
evolution is guaranteed to remain feasible. 

Later, these methods were mostly considered as inefficient until 
Karmakar  triggered a revolution in the field of
optimization by providing the second polynomial time algorithm for linear
programming \cite{khachiyan1980polynomial,karmarkar1984new}. It has
then been realized that Karmakar's algorithm is equivalent to a
particular interior point method
\cite{gill1986projected,bayer1991karmarkar}. Notice that, historically,
the first 
polynomial time algorithm for linear programming is due to 
Khachiyan and  is based on ellipsoid method (which is not an interior
point method). 

 A very elegant presentation
of Karmakar's algorithm and the associated flow, inspired from
\cite{anstreicher1988linear}, can be be found in
\cite{moore2011nature}, including an elegant presentation of its
polynomiality based on ordinary differential equation arguments. Refer
to \cite{renegar2001mathematical} for a general introduction to
interior point methods.





\subsection{Distributed Computing }{\QUID{Olivier}}
\label{sec:populationprotocol}

Recent years have also seen the birth of new classes of models in
\emph{distributed computing}. In particular, the model of \emph{population protocols}
consists of passively
mobile anonymous agents, with finitely many states, that interact in
pairs according to some rules, i.e. a given program \cite{AspnesADFP2004}. \emph{Passively
  mobile} 
means that agents have no control over the other agents with whom
they will interact. The model was initially introduced in the context
of sensor networks, but it is nowadays considered as a fundamental
model for large passively mobile
populations of agents with resource-limited anonymous mobile
agents. 


Most works on the model have considered these protocols as
computing predicates over multisets of states. Given some input
configuration, the agents have to decide whether this input satisfies
a given predicate: the population of agents has to eventually \emph{stabilize} to
a configuration in which every agent is in a particular accepting
(respectively rejecting) state if and only if the predicate is true
(resp. false). The model is \emph{uniform}: the program is assumed to be
independent of the size of the population. 

The seminal work of Angluin \emph{et al.}
\cite{angluin2007cpp,AspnesADFP2004} proved that predicates decided by
population protocols are precisely those on counts of agents definable
by a first-order formula in Presburger arithmetic -- equivalently, this corresponds to semilinear sets.
An elegant proof of this result can be found in \cite{esparza2016verification}.  Note that this
computational power is rather restricted, as multiplication for
example is not expressible in Presburger arithmetic.

The model has been intensively investigated since its introduction.
Several variants have been studied in order to
strengthen it with additional realistic and
implementable assumptions. This includes natural restrictions, like
modifying hypotheses on interactions between agents  (e.g., one-way
communications \cite{angluin2007cpp}, particular interaction graphs
\cite{AngluinACFJP2005}).  This also includes probabilistic
population protocols that assumes random
interactions~\cite{AspnesADFP2004}.  Fault tolerance has
also been considered
\cite{Delporte-GalletFGR06}, including self-stabilizing solutions
\cite{angluin2008self}. For some introductory texts
about population protocols, see for instance \cite{PopProtocolsEATCS,michail2011new}.

The model can be seen as a particular case of the (stochastic)
chemical reaction networks discussed in Section \ref{sec:SCRN}. 

Here, we focus on models and results in the context of distributed
computing, and mainly discuss computability issues. Covering all
works devoted to variants of population protocols in the distributed
computing community is out of the ambition of this chapter: see
\cite{PopProtocolsEATCS,michail2011new} for surveys.
Among many variants of population protocols, the \emph{passively mobile
(logarithmic space) machine model} was introduced by Chatzigiannakis \emph{et
  al.}  \cite{chatzigiannakis2011passively}. It this model, each agent carries a bounded space Turing machine,
  instead of a finite state automaton.  \PLUSCOURT{An exact characterization
\cite{chatzigiannakis2011passively} of computable predicates has been
established: this model computes  $SNSPACE(n S(n)$ as long as $S (n) = \Omega (\log n)$, where
$SNSPACE(n S(n))$ denotes all symmetric predicates in $NSPACE(n S
(n))$, $n$ is the number
of agents, and $O(S(n))$-space Turing machines are considered.  Chatzigiannakis \emph{et al.} \cite{chatzigiannakis2011passively} also
establish that  the
model with  $S(n)=o(\log \log n)$ space per agent is equivalent to population protocols, i.e., to the case
$S(n)=O(1)$. }
In an orthogonal way, \emph{community protocols}, 
where each agent has a unique
identifier and  can only store $O(1)$ other agent
identifiers, exclusively from agents that it met, were introduced by
Guerraoui and Ruppert~\cite{guerraoui2009names}. They proved, using results about the
so-called \emph{storage modification machines} \cite{schonhage1980storage},
that such protocols 
simulate  Turing machines very efficiently\PLUSCOURT{:  the predicates decided by this
model with $n$ agents are precisely the predicates in $NSPACE(n \log n)$.}.
A hierarchy between the two models has also been studied recently, by
considering the case of homonyms, that is to say when several
agents may share the same identifier
\cite{NETYS2015,TOCS16}\PLUSCOURT{: by varying the number of
identifiers, one goes from the case of no identifier corresponding to
population protocol model to the case of unique identifiers
corresponding to the community protocol model. Lower and upper bounds
on the computational power of the model have been derived for several
intermediate cases}.

The population protocols can also capture natural models of
dynamics of some opinion spreading models by considering probabilistic
rules of interactions: results on the convergence and threshold
properties of the so-called Lotka-Volterra population protocol have
been established \cite{czyzowicz2015convergence}.

Notice that many models coming from dynamics of rational agents in the
context of (learning equilibria in) Game Theory can also be
considered as analog distributed models of computation \cite{LivreWeibull,Evolutionary1}.

\subsubsection{Large Population Protocols}
\label{sec:lpp}

When considering probabilistic interaction rules, as in the previously mentioned
settings, the underlying dynamical system is a Markov chain.

If the
population of agents is large, the random process converges to its
(deterministic) limit continuous dynamic given by some ordinary
differential equation (also called its \emph{mean-field limit}): this corresponds to the differential semantics
discussed in Section \ref{sec:limite} for chemical reactions.

These considerations led to the so-called \emph{Large Population Protocols}
\cite{MFCS12}: real numbers which correspond to limit ratio of
programmable dynamics by such model have been demonstrated to
correspond precisely to algebraic numbers \cite{MFCS12}. This results
has many similarities with \cite{huang2017real} obtained in the
context of stochastic reaction networks, but with a slightly different
notion of computability.

A framework to translate certain subclasses of differential equations into protocols for distributed systems has been proposed
\cite{gupta2007design}. 
This is illustrated on several examples either taken from
distributed problematics (responsibility migration, majority
selection) or from classical models of populations such as
Lotka-Volterra model of competition.

\subsection{Black Hole Models }{\QUID{Amaury}}
\label{sec:blackhole}

\AFAIRERELIRE{Quelqu'un qui pourrait relire?}


\emph{Black Hole computations} usually refers to the study of the validity of the Physical
Church Thesis (see Section~\ref{sec:church_thesis}) in the context of Einstein's
General Relativity (GR). In other words, does GR allow for spacetimes where an observer
can observe \text{in finite time} an eternity of some other device. Informally, the
setup is as follows: some device (that we refer to as the computer) will try to
solve some hard problem, like checking the consistency of ZFC. As soon as the computer
finds a counter-example, it sends a signal to the observer, otherwise it keeps
checking ZFC for eternity (since there is an infinite number of formulas to check).
In parallel, an observer will manage to view the result
of this infinite computation in finite time. More precisely, the observer will have a
spacetime location $q$, at which it can check for the existence of the signal: if it receives
a signal, then ZFC is inconsistent, otherwise ZFC is consistent. Hogarth proved
that such spacetimes, usually referred to as \emph{Malament-Hogarth} (\emph{MH}), can exist in theory \cite{Hogarth92}.
Note that in this setting, it is crucial that $q$ is known to the observer,
so that after it reached $q$, the observer knows whether ZFC is consistent or not,
and can use this information. 

This question has received a lot of attention, especially concerning
the physical realization of such a setup \cite{EN93,HogarthPSA94,HogarthThesis,EN02,HogarthTrouNoir06}. It has since emerged that \emph{slowly rotating
Kerr black holes}, and possibly \emph{Reissner-Nordstr\"{o}m} (\emph{RN}) black holes, provide
a plausible physical realization for these experiments. Two very good surveys on the problems and solutions
to the many obstacles encountered so far have been published \cite{NA06,ND06}. We will mention some of these in the
remaining of this section. Another important question is to characterize the extent
of hypercomputation available in a MH spacetime. Hogarth originally showed that any
$\Sigma_1$ set could be decided in the Kerr spacetimes. Recently, it was shown
that MH (but not necessarily Kerr) spacetimes can decide all hyperarithmetic predicates
on integers, but not more (under some assumptions) \cite{Welch06}.

Without giving too many details, we now mention some aspects of the physical
realization of black holes computation. An historical objection was that a \emph{Schwarzschild
black hole} (non-rotating and non-charged) has a punctual singularity to which any
observer is attracted (in this setting, the observer is the one ``jumping'' into the black hole whereas the
computer stays outside), and eventually gets crushed by the ever-growing tidal forces
that tear it appart. However, both slowly rotating Kerr and RN black holes
have negligible tidal forces. Furthermore, their singularity has a shape of ring
that the observer can avoid forever, meaning that the observer could survive infinitely
long within the black hole. In particular, astronomical evidence suggest that such
Kerr black holes exist and have a size roughly that of a solar system. Another
kind of objection was that this setup neglected quantum effects that do not exist
in pure GR, but many of these objections have been solved, although the details
depend on whether the universe is expanding or not.
To summarize with a slightly exaggerated statement, the entire earth population
could jump into a black hole, and continue living inside forever
but now knowing whether ZFC is consistent or not.
Other spacetimes, such as the \emph{anti de Sitter},
theoretically allow to exchange the role of the observer and the computer, meaning that
the observer could stay out of the black hole and send the computer into the black
hole, and get an answer in finite time.

Finally, we mention some related work on \emph{Closed Timelike Curves} (\emph{CTC}s) that, if
they exist, would allow to solve PSPACE problems very quickly but do not allow
for hypercomputation \cite{AaronsonW09}.
Another line of research is to build a logical axiomatization of spacetime, that
is, building relativity theory as a theory in the sense of first-order logic.
For a survey on the subject, see \cite{Andreka2006}.


\subsection{Spatial Models}

We now consider various \emph{spatial} models, that is to say various
models which share a concept of computation based on the distribution
of entities among some space and communications between these
entities. We agree that our classification is debatable: some of the
previous models can be already considered as such (see
e.g. reaction-diffusion systems), or that some of the considered
models can also be considered as continuous space (see
e.g. cellular-automata that can be considered as acting over Baire's
space $\{0,1\}^\N$ homeomorphic to $\R$).

\feedbackOK{Anonymous referee 1}{
Major remarks:

p.11, Section 2.10: I suggest that a short paragraph should be added describing spatial models, before
Section 2.10.1.

\textcolor{red}{OLIVIER}
}

\subsubsection{Computational Fields }{\QUID{Amaury, premier jet d'Olivier}}

\label{sec:computationalfields}

\emph{Computational fields} are analog models of spatial massive parallelism. There
are motivated by the hypothesis that in various contexts it now makes
sense to consider that the number of processing elements is so large
that it may conveniently be considered as a continuous quantity
\cite{maclennan1999field,maclennan1990fieldtheory}.  The theoretical
functions of computational fields, based on tools from functional
analysis, have been studied
\cite{maclennan1990fieldtheory}. Applications of Computational fields
have been explored and advocated in several articles
\cite{maclennan2009analog,maclennan1999field,maclennan2014promise}.

\subsubsection{Spatial Computing Languages }{\QUID{Olivier}}

Previous computational fields models fall naturally in the more
general framework of \emph{spatial computing}: computation has become
cheap enough and powerful enough that a large number of computing
devices can now be embedded into many environment. As a result, a whole
spectrum of \emph{Domain Specific Languages} (DSLs) have emerged to widen the
gap between the application needs of users in various domains
(e.g. biology, reconfigurable computing), usually at a global level,
and the programming,  usually at a local level,  of the the increasingly complex
systems of interacting computing devices.  A common pattern in all those
models and languages is the close relationship
between the computation and the arrangement of the computing devices
in space. For a survey and references, with discussions about many related aspects, see \cite{bealorganizing}.







\subsubsection{Cellular Automata Based Models }{\QUID{Amaury}}
\label{sec:durand}

\emph{Cellular automata} correspond to a particular class of spatial
computing models. The classical model is an abstract discrete time and
space model, with a rather well studied computation theory. We
will focus here on developments related to analog models. \TODO{on
  pourrait parler de asynchronous.}


There exist several Cellular Automata (CA) inspired models, usually built as a 
abstraction of CA in the limit where cells are infinitely small.

One such model is based on an infinite tessellation of space-time \cite{schaller2009scale}. It has been
compared to classical models from computability theory and
proven to be capable of hypercomputations  \cite{schaller2009scale}. 

Smooth (continuous) versions of game of life have been investigated
\cite{hiebeler2006dynamics}, yielding very elegant dynamics.
More generally, the passage from cellular automata to continuous
dynamics (partial differential equations) has also been investigated
in \cite{cervelle2013constructing}.

Another model with several recent developments is that of \emph{Signal Machines} (SM) introduced in \cite{Durand-Lose05}
where dimensionless signals are synchronously moving on a continuous space, and local update rules
are used to resolve collisions. The
major difference with CA is that both time and space are continuous (\emph{i.e.} reals
instead of integers). A SM configuration is given by a partial mapping from $\R$ (the
space) to a finite number set of \emph{meta-signals}, essentially each signal has
a type. Each SM comes with a finite set of \emph{collision rules}. When
two or more signals meet, a \emph{collision} happens, all incoming
signals are destroyed and the rule gives a list of outgoing signals that are created.
Between collisions, each signal moves in a direction at a certain speed that depends
on its type. Thus there are only finitely many different speeds. One can think of
the speed of signals as the \emph{slope} of its line in the 2D space-time diagram.
Figure~\ref{fig:signal_machines}, illustrates this
process on a simple example. In a series of papers (\cite{DurandLose06} and onwards),
Durand-Lose and coauthors investigate the computation power of this model with various
restrictions. It is possible to embed Black Hole computations, Turing machines, BSS
and more in this model, depending on whether irrational numbers and accumulation
points are allowed. Since the unrestricted model exhibit Zeno phenomenon, it is
super-Turing powerful.

\begin{figure}
    \begin{minipage}[b]{.33\linewidth}
        \begin{subfigure}[b]{\linewidth}
        \centering\begin{tabular}{c|c}
        \textbf{Name}&\textbf{Speed}\\
        \hline
        Add, Sub&1/3\\
        A, E&1\\
        O, W&0\\
        $\overrightarrow{\text{R}}$&3\\
        $\overleftarrow{\text{R}}$&-3\\
        \end{tabular}
        \caption{meta-signals}
        \end{subfigure}\\

        \bigskip
        \begin{subfigure}[b]{\linewidth}
        \centering\begin{tabular}{rcl}
        \{ Add, W \} & $\rightarrow$ & \{ W, A, $\overrightarrow{\text{R}}$ \}\\
        \{ $\overrightarrow{\text{R}}$, W \} & $\rightarrow$ & \{ $\overleftarrow{\text{R}}$, W \}\\
        \{ A, $\overleftarrow{\text{R}}$  \} & $\rightarrow$ & \{ O \}\\
        \{ $\overrightarrow{\text{R}}$, O \} & $\rightarrow$ & \{ $\overleftarrow{\text{R}}$, O \}\\
        \{ Sub, W \} & $\rightarrow$ & \{ W, E \}\\
        \{ E, O \} & $\rightarrow$ & \{ \}
        \end{tabular}
        \caption{collision rules}
        \end{subfigure}%
    \end{minipage}%
    \begin{subfigure}[b]{.66\linewidth}
    \centering\begin{tikzpicture}
        \draw[red,thick,dashed] (-1.5,0) -- (0,4.5) node[left,midway,black] {Sub}
            -- (0.75,5.25) node[above,midway,black] {E};
        \draw[darkgreen,thick] (-1,0) -- (0,3) node[left,midway,black] {Add};
        \draw[darkgreen,thick] (-1/6,0) -- (0,0.5) node[left,midway,black] {Add};
        \draw[very thick] (0,0) -- (0,0.5) node[right,midway] {W}
            -- (0,3) node[right,midway] {W} -- (0,4.5) node[right,pos=0.7] {W}
            -- (0,5.4) node[left,midway] {W};
        \draw[thick,blue,dotted] (0,0.5) -- (3,1.5) node[midway,below,black] {$\overrightarrow{\text{R}}$}
            -- (1.5,2) node[midway,above,black] {$\overleftarrow{\text{R}}$};
        \draw[thick,darkgreen] (0,0.5) -- (1.5,2) node[midway,above,black] {A};
        \draw[thick,blue,dotted] (0,3) -- (1.5,3.5) node[midway,below,black] {$\overrightarrow{\text{R}}$}
            -- (0.75,3.75) node[midway,above,black] {$\overleftarrow{\text{R}}$};
        \draw[thick,darkgreen] (0,3) -- (0.75,3.75) node[midway,above,black] {A};
        \draw[thick] (0.75,3.75) -- (0.75,5.25) node[right,midway] {O};
        \draw[thick] (1.5,2) -- (1.5,3.5) node[right,midway] {O}
            -- (1.5,5.4) node[right,midway] {O};
        \draw[very thick] (3,0) -- (3,1.5) node[right,midway] {W}
            -- (3,5.4) node[right,midway] {W};
        \foreach \x/\y in {0/0.5, 0/3, 0/4.5, 0.75/3.75, 0.75/5.25,
                1.5/2, 1.5/3.5, 3/1.5} {
            \draw[inner sep=1pt,black,fill=white] (\x,\y) circle (1.5pt);
        }
        \draw[<->] (-2.2, 0) -- (3.2, 0) node[right] {$\R$} node[below,midway] {Space};
        \draw[->] (-2, 0) -- (-2,5.5) node[above,left] {$\Rp$} node[left,midway,rotate=90,yshift=0.7em,xshift=2em] {Time};
    \end{tikzpicture}
    \caption{space-time diagram}
\end{subfigure}
\caption{Signal machine that finds the middle between two input vertical signals.
    The process is started by the arrival of the Add signal from the left. It triggers
    a sequence of collisions that results in a vertical signal O that is positioned exactly
    halfway between the two W signals. It is possible to generate the middle between
    the left W and O signal by sending another Add signal. It is also possible to
    suppress the leftmost O signal by sending a Sub signal from the left. Finding
    the middle is a key primitive in SM computations.}
\label{fig:signal_machines}
\end{figure}





\subsection{Other Various Models}{\QUID{Olivier}}
\label{sec:reaction-diffusion}

\feedbackOK{Erik Winfree}{
Some nice papers on the computational power of reaction-diffusion systems include:
\cite{scalise2014designing} Scalise, Dominic, and Rebecca Schulman. "Designing modular reaction-diffusion programs for complex pattern formation." Technology 2.01 (2014): 55-66.
\cite{scalise2016emulating} Scalise, Dominic, and Rebecca Schulman. "Emulating cellular automata in chemical reaction-diffusion networks." International Workshop on DNA-Based Computers. Springe, 2014.
}

\feedbackOK{Erik Winfree}{
Papers on computational aspects of molecular self-assembly models include:
\cite{evans2017physical} Evans, Constantine G., and Erik Winfree. "Physical principles for DNA tile self-assembly." Chemical Society Reviews (2017).
\cite{soloveichik2007complexity} Soloveichik, David, and Erik Winfree. "Complexity of self-assembled shapes." SIAM Journal on Computing 36.6 (2007): 1544-1569.
}

\feedbackOK{Andrew Adamsky}{ 
BTW, you might consider reaction-diffusion computing, see PDF of the book [Andy Adamatzky, Ben De Lacy Costello and Tetsuya Asai.
Reaction-Diffusion Computers. Elsevier, 2005. ISBN: 978-0-444-52042-5]
\url{
https://drive.google.com/open?id=0BzPSgPF_2eyUVmVvR3N3MHhVLTQ}

Reference: \cite{adamatzky2005reaction}
Obviously much more was published between 2005 and now, but I will not bother you with tons of papers.
\textcolor{red}{OLIVIER}
}

\feedbackOK{Andrew Adamsky}
{Good evening,

have been think about your chapter. In principle the following cases of --- implemented in experimental laboratory conditions --- unconventional computers can be classed as spatial

1. based on propagation and interaction of waves (reaction-diffusion computers, soliton-based computers)

2. based on propagation of patterns/solid/soft forms (slime mould computers, plant root computers, crystallisation computers)

3. based on propagation of swarms of creatures (e.g. crab gates)

In case you are interested to read something about slime mould here are some books

\url{
https://drive.google.com/open?id=0BzPSgPF_2eyUX21TdnRkN0VWejQ}

Reference: \cite{andrew2010physarum}

\url{
https://drive.google.com/open?id=0BzPSgPF_2eyUQ3J1MXVSVnAweWc
}

Reference?

\url{
https://drive.google.com/open?id=0BzPSgPF_2eyUZDhxUGhUTWZWODA
}

Reference?

I might email to you again if have more ideas.

Best wishes,
Andy

\textcolor{red}{OLIVIER A BAFFER} } 

We do not intend to cover in this
document all models. Various other unconventional computers can also
be classified as spatial. This  includes models based on propagation
and interaction of waves (reaction-diffusion computers \cite{scalise2016emulating,scalise2014designing,adamatzky2005reaction}, soliton-based
computers). Or models based on propagation of pattern 
forms (slime mould computer \cite{andrew2010physarum}, plant root computers, crystallisation
computers). Or models based on propagation of swarms of creatures
(e.g. crab gates). This also include molecular self-assembly models
\cite{evans2017physical, soloveichik2007complexity}.


\AFAIRERELIREE{Milici}

The frontier between analog and non-analog models of computation is
not so clear, and some models are clearly at the frontier. This
includes the optical models of \cite{Woods05}, or some quantum
computation models such as \cite{Deu85,Gru97,Sho94,Kieu04}, or
finite 2-dimensional coupled map lattices that have been proven to be
computationally universal \cite{orponen1996universal}.

\emph{Planar mechanisms} were claimed to correspond to all finite algebraic curves by Kempe
\cite{Kempe76}. The initial statements of Kempe had flaws that were
corrected and extended in \cite{Smi98}.  \emph{Tractional Motion machines}
constitute an extension of this models with a rich and elegant theory
discussed in \cite{DBLP:journals/ijuc/Milici12,
  DBLP:conf/uc/Milici12}. The fact that some functions computable in
this model are not GPAC computable has been established
\cite{DBLP:journals/ijuc/Milici12}.  Some discussions on limitations of machines based on
the historic concept of compass and ruler constructions can be found in \cite{MCC06}.



\section{Dynamical Systems and Computations}

\subsection{Arbitrary vs Rational/Computable
  Reals}{\QUID{Amaury}}
\label{sec:unrela}




Before delving into the core
of the topic of \emph{dynamic undecidability}, we would like to point out
a fundamental difference between discrete and continuous dynamical
systems: \emph{real numbers}. Indeed, it is clear that many objects involved in
discrete computations (integers, rational numbers, graphs, etc) have a
finite representation. For example, one can represent an integer in
base 2 by its bits. This property is no longer true for real numbers,
independently of the details of the representation. In fact, for
reasonable representations (see \cite{Wei00} for more on the theory of representations) of the real
numbers, a \emph{single real number} can turn out to be a very
powerful object. One such example is the real number, sometimes called the
\emph{``Turing number''}, whose $n^{th}$ bit is 1 if and only if the $n^{th}$
Turing machine halts\footnote{One needs to fix a particular
  enumeration of Turing machines.}.  Intuitively, any powerful enough
dynamical system that has access to this number can solve the Halting
problem for Turing machines in finite time, and is thus super-Turing
powerful. Other numbers, such that $\Omega$ of Chaitin
\cite{chaitin2007halting}, can turn out to be even more powerful. A
consequence of this fact is that dynamical systems are able to solve
(classically considered)  uncomputable problems 
 if one allows such numbers to appear in the
system. This is why it is often important to restrict to systems with
computable or rational descriptions if one does not want to
consider models with some hypercomputational features.  

Understanding the obtained computability theory, if this hypothesis is
not done, i.e. when arbitrary (possibly non-computable) reals are
allowed in the description of systems, gave
birth to a whole set of developments discussed in Section
\ref{sec:nn} and \ref{sec:physical}.

In the rest of this section, in order to avoid to consider such classes of models,
we assume that we restrict to rational or computable descriptions. 

\subsection{Static  Undecidability}{\QUID{Amaury}}

\AFAIRERELIREE{Keijo Ruohonen}
It is, however, often rather easy to get
undecidability results about dynamical systems by observing that
integers (or discrete sets) can be considered as particular reals, and
hence that all undecidability problems known over the integers have some counterpart
over the continuum \cite{Ruo97}. This allows to map well known undecidability
questions of recursive analysis (e.g. testing whether a real is zero)
or of computability theory (e.g. testing whether a given polynomial
with integer coefficients has an integer root) into the setting of
dynamical systems. This is sometimes known as \emph{static
  undecidability}. For concrete simple examples of results obtained with that spirit, see \cite{Asa95,Ruo97}.
More elaborate examples are
the proof that determining whether a polynomial dynamical system has a
Hopf bifurcation \cite{costa1994undecidable} or is
chaotic \cite{costa1993dynamical} is undecidable.

\subsection{Dynamic Undecidability 
}{\QUID{Amaury}}\label{sec:undec_dyn_sys}

\feedbackNIOK{Emmanuel Jeandel}{ Salut,
(c) Dans le discours de 3.3, peut-être que vous pouvez citer Collins
et van Schuppen (Observability of Hybrid Systems and Turing machines)
qui donne quelques détails utiles (dont la Shadowing Property)

Reference = \cite{ColSch04a}

\textcolor{red}{OLIVIER}
}

\emph{Dynamic undecidability}, as opposed to \emph{static
  undecidability} \cite{Ruo97}, corresponds to a proof technique
that is often used with various models of dynamical systems. In this
case, the undecidability is obtained by proving that for every
Turing machine, one can build a dynamical system in the considered
class able to simulate step by step the Turing machine. 

 To be more concrete,
let us
informally describe an embedding of Turing machines into dynamical systems with
discrete time and continuous space.
We can encode the configuration of a Turing machine using a vector
$(l,q,r)$ where $l,r\in[0,1]$ will encode the tape (left and right side of the tape),
and $q\in[0,1]$ will encode the state, for example the $i^{th}$ state is mapped to $\frac{i}{N}$
where $N$ is the number of states. To encode the tape, we simply view the content of
the tape as a list of digits. In the case of a binary alphabet, a word $w\in\{0,1\}^*$
will be encoded by the real $0.w=\sum_{i=1}^{|w|}w_i2^{-i}$. By convention, we will
say that the symbol under the head is the first symbol on the right tape.

To simulate one step of the machine, we need to perform tests on the state
and then apply the corresponding action. To find in which state we are, we need
comparisons of the form $q=^?\tfrac{i}{N}$. To find which symbol is
under the head (the first
symbol of the right tape $r$), note that if $r_1=0$ then $0.r<\tfrac{1}{2}$,
but if $r_1=1$ then $0.r\geqslant\tfrac{1}{2}$. Thus we only need a test of the
form $r<\tfrac{1}{2}$. To move the head of the machine, we will only need addition
and multiplication by a constant. For example, if we want to move the tape to the
right, then we perform the assignment $r\leftarrow\tfrac{\sigma+r}{2}$ where
$\sigma$ is the symbol written by the head, and $l\leftarrow 2(l-\rho)$ where
$\rho$ is the first symbol of the left tape. Note that we can easily have access
to $\rho$ the same way we have access to $\sigma$ and we can make a case distinction
between $\rho=0$ and $\rho=1$. Finally we assign the new state with $q\leftarrow\tfrac{i'}{N}$
where $i'$ is the new state.

To summarize, in order to simulate a Turing machine with a discrete time and
continuous space machine, we need comparisons against fixed values
($=x$ and $<x$), assignment, addition and multiplication by a
constant. Importantly, all the constants involved are rational
numbers.  One may be worried that exact comparisons between real
values is an unrealistic assumption, there are two answers to this
issue. First, analog models of computation are not necessarily
concerned with realistic assumptions that come from the world
of discrete computations (or  from computable analysis's basic assumptions about computability): comparing
two real numbers is a very natural operation in several continuous
models of computations. 

Second, and more importantly, even if exact tests are not possible,
this simulation does not really require exact tests. Without going
into too many details, notice that on the state, there is gap of size
$\tfrac{1}{N}$ between the encoding of two states, thus an approximate
test will suffice. Similarly for the tape, a classical trick is to
encode a binary tape in base $4$, by $0.w=\sum_{i=1}^{|w|}2w_i4^{-i}$.
This way, if $r_1=0$ then $0.w\leqslant\tfrac{1}{6}$ (the worst case
is $0.02222\ldots$) but if $r_1=1$ then $r\geqslant\tfrac{1}{2}$,
leaving again a constant gap between the two options. 

All the above settings provides the basic ingredients for the possibility to simulate
any Turing machine by a discrete time model. Similar \emph{static
undecidability} constructions have been used in various models:
general dynamical systems \cite{Moo90,Ruo93,Bra95,Ruo97,ColSch04a}, piecewise affine maps \cite{KCG94},
sigmoidal neural nets \cite{SS95}, closed form analytic
maps \cite{KM99}, which can be extended to be robust \cite{dsg05}, and
one dimensional restricted piecewise defined maps \cite{Potapov05}.

If one wants to simulate a Turing machine by a continuous time model,
then several techniques may be used. One is to build a continuous time
system such that the crossing of the dynamics with a given
hyperplane  (its \emph{Poincaré's map}) corresponds to a discrete
time system as above.  Another one is to build a continuous time
system such that its projections at discrete time $t=0,1,\dots$ (its
\emph{stroboscopic view})  is given by a discrete
time system as above.

\feedbackNIOK{Manuel Lameiras Campagnolo}{
I wonder if the $\theta_\infty$  version of the clocks on top of page
20 really comes from Branicky95: I did use that function in my PhD
thesis  (after definition 2.4.8 and in Proposition 3.4.1) but perhaps
I didn't realize it had been proposed earlier; 
\textcolor{red}{OLIVIER ANCIEN SURVEY}
}

Usual such constructions rely on a suitable generalization of the idea
of \emph{``continuous clocks''} \cite{Bra95} for iterating
functions over the integers. For simplicity, we will use a reformulation
of these original equations, based on \cite{Campagnolo}.
The idea is to start from the function
$f: \R \to \R$, preserving the integers, and build the ordinary
differential equation over $\R^3$ 
$$
\begin{array}{lll} 
y_1' &=  & c(f(r(y_2))-y_1)^3 \theta(\sin (2\pi y_3)) \\
y_2' &= & c(r(y_1) - y_2)^3 \theta(-\sin(2\pi y_3)) \\
y_3' & = & 1. \\
\end{array}
$$
where $r(x)$ is a rounding-like function that has value $n$ whenever
$x \in [n-1/4,n+1/4]$ for some integer $n$, $\theta(x)$ is $0$ for
$x \leq 0$ and $\exp(-1/x)$ for $x>0$, and $c$ is some suitable constant. A simple analysis of this dynamics shows
that it basically alternates $y_1\leftarrow f(y_2)$ and $y_2\leftarrow y_1$. As a result, the
stroboscopic view of this differential equation is the discrete-time system $y(n+1)=f(y(n))$.

Indeed \emph{dynamic undecidability} constructions have been used in
various continuous time models, including piecewise constant maps  \cite{AMP95},
general dynamical systems \cite{Bra95} and  polynomial ordinary differential equations \cite{dsg05}.






Observe that dynamic undecidability has merits not covered by
static undecidability.
Considering Turing machines as dynamical systems provides a view not
covered  by the von Neumann
  picture \cite{Campagnolo}.  This also shows that many qualitative
  features of (analog or non-analog) dynamical systems, e.g. questions
  about basins of attraction, chaotic behavior or even periodicity,
  are not computable \cite{Moo90} even for a fixed dynamical system
  (corresponding for example  to a universal Turing machine). 
Conversely, this brings into the
  realm of Turing machines and computability general questions
  traditionally related to dynamical systems \cite{CIEChapter2007}. These include, in
  particular, the relations between universality and chaos
  \cite{Asa95}, necessary conditions for universality
  \cite{DelKurBlo04}, computability of entropy \cite{Koi01} and
  understanding of edge of chaos \cite{LM05}. 



One may object that the above simulations require unrealistic hypotheses on
systems. For example, that it requires the ability of doing
computations with arbitrary precisions: $O(2^{-n})$ precision
corresponds roughly to size $O(n)$ of the tape in the simulation
above. But at this point, in order not to overinterpret such statements,
we believe that some digression about computational model would be clarifying.
This is the topic of the next section.

\section{Philosophical, Mathematical and Physics Related Aspects }

\subsection{Mathematical Models vs Systems }{\QUID{Olivier}} 

\label{sec:limite}

While the use of dynamical systems is very
common in experimental sciences (biology, physics, chemistry, etc...),
it is important to remember that this is an abstract (mathematical)
view, and that the properties of those systems can actually be different in
many aspects from those of the initial system they intend to model.

In the literature, there are many accounts in various contexts
about situations where models behave very differently
according to the levels of modelization. As an illustrative
example, we can mention the Lotka Volterra equations
discussed in \cite{Les06,KLT06}. It is
shown in these articles that no classical ordinary differential
equation method (unless biased on purpose) behave as stated in
classical studies of these equations: no closed cyclic behaviour is
observed for example, and ``energy'' like functions supposed to be
preserved are not.  While the continuous Lotka Volterra dynamic is
usually presented as an abstraction of a discrete setting, these
results can be seen as a nonmatching of behaviours between
discrete and continuous models. 








A context where various semantics   
have been discussed, with clear stated vocabulary, is that
of chemical reactions. To a given formal set of chemical reactions, a hierarchy of semantics
can be associated at different levels of abstraction
\cite{fages2014cells}. 

The most concrete (low-level) interpretation is provided by
the \emph{Chemical Master Equation (CME)} which defines the
probability of being in a state $x$ at time $t$ by considering the
system as a continuous time Markov chain governed by \emph{propensity} of
reactions over discrete number of molecules, giving the rate of
reactions of the associated chain. 
The \emph{differential semantics} describes the evolution of the
system by an ordinary differential equation (ODE). This ODE can be
seen as the mean-field view of the reactions, and can be considered as
being derived from the CME by a first-order approximation.
Then the \emph{stochastic semantics} is defined by
considering a Continuous Time Markov Chain (CTMC) over integer numbers
of molecules (discrete concentration levels).  The very classical
algorithm of Gillespie \cite{Gil77} provides a method to simulate
efficiently this CTMC from the reactions.
If the simulation provided by this algorithm is often 
similar to the ODE simulation for large number of molecules, it may
exhibit qualitatively different behaviors, in particular when the
number of molecules is small \cite{fages2014cells}.

The abstraction of stochastic semantics by simply forgetting the
probabilities also yields to the \emph{Petri net semantics} of the
reactions, where the discrete states define the number of tokens in
each place, and the transitions consume the reactant tokens and
produce the product tokens.
The abstraction of the Petri net semantics
into \emph{Boolean semantics} is then obtained by reasoning only on
the absence/presence of a given molecule.

All these discrete and stochastic trace semantics of reactions
systems can be related by a Galois connection in the framework of
abstract interpretation \cite{fages2008abstract}. However, it is
important to alway remember that the behavior of a given semantics can
differ from the others in many ways.

The study of the relationship between the differential and
the stochastic semantics dates back to the seminal work of
Boltzmann in the 19th century who created the domain of Statistical
Physics. In this setting, the differential semantics is obtained from
the stochastic semantics by limit operations, where the number of
molecules tends to infinity and the time steps to zero, under
several assumptions such as perfect diffusion
\cite{fages2008abstract}.  
More generally, there is a whole mathematical theory
justifying the passage from a stochastic dynamic over a (usually huge)
population of agents to its first-order ordinary differential equation
description \cite{kurtz1981approximation}. However,
hypotheses of mathematical theorems are not always valid in the
context of experimental science applications and concerns are
sometimes very different. Notice that the specific
context of chemistry has been discussed in articles such as
\cite{gillespie1992rdc}.

All these discussions emphasize the difference between a system and
its models, and even between the various levels of abstraction of a
given system, and the importance of stating explicitely the considered
semantics in the coming discussions.

\subsection{Church Turing Thesis and Variants }{\QUID{Olivier}}\label{sec:church_thesis}

A common statement of the Church Turing Thesis is that \emph{every
  effective computation can be carried out by a Turing machine, and
  vice versa.}
However, it is often misunderstood, and an instructive discussion  
demonstrating the improper semantic shift from its initial
statement and the way it is often misunderstood today can
be found in \cite{ChurchThesis}.  We repeat some of these elements here.

\feedbackOK{Anonymous Referee 2}
{

- Page 22, line 8 of section 4.2: the expression “but terms of art of
these disciplines” seems confusing.

Ceci étant, c'est me semble t'il (à vérifier) exactmenet comment ca
que c'st dit dans l'article cité.
\textcolor{red}{OLIVIER CITER}
}
\feedbackOK{amaury}{Je comprend pas l'expression "terms of art of these disciplines"}

Following \cite{ChurchThesis}, ``The Church Turing thesis concerns the
concept of an effective or systematic or mechanical method in logic,
mathematics and computer science. ‘Effective’ and its synonyms
‘systematic’ and ‘mechanical’ are terms of art in these disciplines:
they do not carry their everyday meaning.'' 

A myth seems to have arisen progressively in several documents about
the fact that Alan Turing's 1936 paper was establishing facts about
limitations of mechanisms  \cite{ChurchThesis}. The thesis that \emph{whatever can
be calculated by a machine} (working on a finite data in accordance
with a finite program of instructions) \emph{is Turing machine-computable}
(sometimes referred to as the \emph{Physical Church Turing
thesis}, or \emph{Thesis M} of \cite{Gan80}) is a very different statement from
the Church Turing thesis.

Thesis M itself admits two interpretations, according to whether
the phrase \emph{can be generated by a machine} is taken in the sense
\emph{can be generated by a machine that conforms to the physical laws
  (without resource constraints) of the actual world}, or in a wide
sense whether the considered
machine could exist in the actual world \cite{ChurchThesis}. Under the
latter interpretation, the thesis is clearly wrong, and hence
not really interesting:  see all examples of
hypercomputations mentioned in next subsection. 

The discussion above suggest that it may also be important to distinguish
machines from their models. The thesis that \emph{whatever can be calculated
by a given model of machine in model(s) of our physical world is
Turing machine-computable} might still be one more  different statement.
\AFAIRERELIREE{Jack Copeland}


\subsection{Are Analog Systems Capable of Hypercomputations?}{\QUID{Olivier}}

\AFAIRERELIREE{MacLennan}

Several results have shown that analog systems are capable of
hypercomputations.
One classical way to do so is to
consider non-computable reals in the machine (see previous
discussion in Section \ref{sec:unrela} and coming Sections
\ref{sec:nn} and \ref{sec:physical}).  Another classical way is to use Zeno's
phenomenon: the possibility of simulating an unbounded number of
discrete steps in a bounded continuous time (e.g. \cite{AM95, Moo95b,
  Bou97b}, see section \ref{sec:zeno}). A presentation of various systems capable of hypercomputations,
with in particular some accounts on analog models of computations, can be found in \cite{syropoulos2008hypercomputation}.

Such results may give some clues about the fact that the
considered model or systems are indeed very/super powerful. But, as already stated in \cite{maclennan2004natural},
we believe that one of the main interest of such
results is not really a mean to advocate models with
hypercomputational power, but rather a way to demonstrate that an extended
definition of computation and computability theory including
alternative (especially analog) models in addition to the Turing machines
are needed \cite{maclennan2009analog, maclennan2004natural}. 

We also personally believe that all variants of the Church Turing
Thesis above are actually about ``\emph{reasonableness}'': what is a reasonable
notion of mechanical method in mathematics and logic for the Church Turing thesis,
what is a reasonable machine for Thesis M, or what is a reasonable
model of the physics of our world for the last
variant. In a contrapositive way, establishing a hyper-computability
result is a way to outline what should (possibly) be
corrected in the model to make it reasonable \cite{Bou05}. 

Notice that this has
been used for example to advocate for corrections of models from
Physics by Warren D. Smith in several papers \cite{Smi2,Smi06},
taking the Church Turing as a postulate, or if one prefers in the context
of physics, as a physical law of our physical world. 

For accounts on physical limitations against
hypercomputational possibilities, one can refer to \cite{cockshott2008physical}.



\subsection{Can Analog Machines Compute Faster?}{\QUID{Olivier}}
\label{sec:hyper}
\label{sec:zeno}

All previous variants also have effective versions, dealing with
complexity, sometimes called the \emph{Effective Church Turing Thesis} or
\emph{Strong Church Turing Thesis (SCTT)}. We now come to the
question of whether analog machines satisfy these versions. 

In 1986, one of the first paper devoted to the explicit question of
whether analog machines can be more efficient than digital ones was
published \cite{VSD86}. The authors claim that the SCTT is provably
true for continuous time dynamical systems described by any
Lipschitzian ODE $y^\prime=f(y)$.  However, their proof assumes that the time variable remains
bounded. No clear arguments is established for the
general case, and this was considered as an open problem until very
recently. 
Interesting results and discussions on the subject can also be found
in \cite{blakey2010model}. 

The question of whether analog machines satisfy SCTT turns out to be
deeply related to the question on how ressources such as time is
measured. In short, the difficulty is that the naive idea of using the
time variable of the ODE as a measure of ``time complexity'' is
problematic, since time can be arbitrarily contracted in a continuous
system due to the ``Zeno phenomenon''. For example, consider a
continuous system defined by an ODE $ y^{\prime}=f(y)$ where
$f:\R\to\R$ and $y:\R\to\R$. Now consider the system
\[
\left\{
\begin{array}
[c]{l}%
z^{\prime}=f(z)u\\
u^{\prime}=u
\end{array}
\right.
\]
where $u,z:\R\to\R$. It is not difficult to see that this
system rescales the time variable and that its solution
is given by $u(t)=e^t$ and
$z(t)=y(e^t)$. Therefore, the second ODE simulates the first, with an exponential acceleration.

In a similar manner, it is also possible to present an ODE which has a
solution with a component $u:[0,\tfrac{\pi}{2})\to\R$ such that
$u(t) = y(\tan(t))$, i.e.~it is possible to contract the
whole real line into a bounded set. Thus any language computable by
the first system (or, in general, by a continuous system) can be
computed by another continuous system in time $O(1)$. This problem has
been observed or used in many continuous models
\cite{Ruo93,Ruo94,Moo95b,Bou97b,Bou99,AD91,CP01,Davies01,Copeland98,Cop02}.

Notice that in addition to time contraction, space contraction is
also possible, by considering changes on space variables. Note however, that
space contractions are more model-dependent and are not always possible. For example, one could
consider the system $z(t)=y(t)e^{-t}$, which makes the system exponentially smaller. However,
doing so with a polynomial differential equation (for example) requires to add a
variable $u(t)=e^{t}$ to the system, thus making the contraction less useful, if useful at all.


Such time or space contraction phenomena seem, in many systems, to physically correspond
to some infinite energy. However, \emph{energy} is not a
mathematically universally defined concept, and defining a robust notion of ``time complexity''
in those systems was an open problem until lately.

\TODO{Discuter quelquepart du chapitre de Graca et al: avec par
  exemple:  1- \cite{Ko83}, \cite{Ko91}, \cite{Smi1}.
2) \cite{MM93}, where it is proved 
using recursive analysis that if $f$ is analytic and 
polynomial time computable then the solution is also polynomial time 
}
It has only recently been shown that analog systems satisfy SCTT,
if ``time'' (as in ``time complexity'') is measured by the length of the trajectory \cite{ICALP2016,JournalACM2017}, and
if considering polynomial ordinary differential equations.  Since this
class of ODEs is very wide and covers in practice most
reasonable classes, this can be considered as a definitive proof of
the statement that many analog systems satisfy SCTT. 

Note that this does not cover all
  variants of the SCTT, since ODEs do not cover all known physics. In particular,
  models that rely on quantum effects, General Relativity, and more generally
  models that rely on physical experiments (see Section~\ref{sec:physical} for example)
  do not fall into this class. In other words, this does not cover the ``physical''
  variants of the SCTT.




\subsection{Some Philosophical Aspects }{\QUID{Olivier}}

\feedbackNIOK{Cameron Beebe}{Thanks for the email, Amaury and Olivier.
You might want to also take a look at Care's 2010 book Technology for Modelling: Electrical Analogies, Engineering Practice, and the Development of Analogue Computing
\textcolor{red}{OLIVIER}
}

\AFAIRERELIREE{Florent Franchette + Mael Pégny + Beebe + Gilles Dowek}

Some of the previously mentioned results 
led to discussions in the context of
philosophy.

First, 
the consideration of systems that could realize hypercomputation gives rise
to the following philosophical question: since the physical construction of a
hypercomputational model is out of the framework of computability
theory, answering the \emph{hypercomputation thesis} requires the
construction of a physical model of hypercomputation. It has been argued that,
even if such a model is built, it would be impossible to verify that
this model is indeed computing a function non-computable by a Turing
machine \cite{Franchette,davis2004myth}. However, it was pointed out that this objection
is not specific to hypercomputation since it is already not possible to verify that a Turing
computable function is correct in finite time in general \cite{Cop04}. Furthermore
simply because there is no systematic way of computing the values of the halting function does
not mean that one could not (in principle) assign one mathematician to each value and ask each of them to
come up with a different method, therefore bypassing the problem of uniformity of computation.

The possibility of new computational models which may be exponentially
more efficient than any previously known machine (in a wide context,
including for example models based on quantum mechanics) suggest
alternative, empirical views of  complexity theory,
usually considered as part of logic and computer science. The
arguments in favor of this view and its consequences have been
discussed in \cite{Peg1304, pegny2013limites}. 
\feedbackOK{amaury}{J'ai reformulé les phrases qui suivent, vérifier qu'elle veulent bien toujours dire la
même chose. \textcolor{red}{OLIVIER}}%
It raises some fundamental questions such as whether the empirical
limits of computing are identical to limits of algorithms or, to put it differently, what is the
the capability of symbolic processes to simulate empirical
processes. This helps to study the epistemology of computations
realized by a machine. In particular, the existence of a function computable by a machine but
not by an algorithm would imply that some mathematical problem are solvable by empirical processes
without any mathematical method \cite{Maellimite,pegny2013limites}.

The statement that there is no equivalent of the Church-Turing thesis for
computation over the reals (often relying on the non convergence of
formalisms as for discrete computations) is shown in
\cite{pegny2016make} to confound the issue of the extension of the
notion of effective computation to the reals on the one hand, and the
simulation of analog computers by Turing machines on the other
hand. It is possible in both
cases to argue in favor of a Church-Turing thesis over the reals \cite{pegny2016make}.
It has also been argued that analog computation literature is often mixing
both the concept of 
continuous valued computations and analog machines, and that a
concept called \emph{model-based computation} can help to untangle the
misconception, by offering a two dimensional view of computation: one
dimension concerns models and the other the type of variables that are
used \cite{Beebe2017}. 

Reflexions on whether nature can be considered as computing, and even
about the notion of computation have also been discussed
\cite{longo2011mathematics} in light of new technologies in
physics and new ideas in computer sciences such as quantum
computing, networks, non-deterministic algorithms. 

Some original views on the Church Turing thesis have also been expressed
by Dowek \cite{dowek2012physical,LivreDowek}. The
statement that all physically realized relation can be expressed by a
proposition of the language of mathematics (that can be called
\emph{Galileo thesis})  can be considered as a consequence of the
physical Church thesis \cite{dowek2012physical}. 

\TODO{Parler ou pas quelqupart de ce que fait Pablo Arrighi's}

\TODO{Mentioner?: Brain computation}

\TODO{ COSTA \url{https://www.researchgate.net/publication/262152565_Computable_Scientists_Uncomputable_World}
Il y aussi toute une littérature historique.... 
}


\section{Theory of Analog Systems}

We now review the various computational theories that have been
developed for analog systems.

\subsection{Generic Formalizations of Analog Computations}

There have been several approaches to formalize
analog computations in a general setting.

\subsubsection{Formalization by Abstract State Machine }{\QUID{Olivier}}

\AFAIRERELIRE{Gurevich?}
\AFAIRERELIREE{ Nachum Dershowitz}

A formalization of generic algorithms covering both analog
and classical, discrete and continuous time algorithms has
been proposed in \cite{BDN16}. This framework, extending
\cite{TAMC12}, is based on an extension and reformulation of
\emph{abstract state machines} (ASM) from Gurevich \cite{Gur00}. The notion
of analog ASM program is introduced, and a completeness result is
proven: any process satisfying the three postulates generalizing those
of \cite{Gur00} is demonstrated to correspond to some analog ASM
program \cite{BDN16}. This is intended to be a first step towards a
formalization and a proof of an equivalent of a Church Turing thesis
for analog continuous and discrete time systems in the spirit of what
has been achieved for discrete-time
models~\cite{boker2008church,dershowitz2008natural,3P}.

\TODO{papier infinitesimals ICALP}

\subsubsection{Formalization by Fixed Point Techniques} 

\feedbackOK{Jeffery Zucker}{

In connection with your remarks on page 26 concerning paper [286]:
You may perhaps want to mention two related papers:

286= je pense erreur

288 = \cite{TZ07}

289 = \cite{TZ11}

(1) N.D. James and J. Zucker (2013):
   A class of contracting stream operators.
   The Computer Journal, 56:15-33.

REFERENCE \cite{james2012class}

   Here it was shown how the results for the fixed point semantics
   of networks developed in [288] can be improved,
   by re-modularizing the network suitably.

(2) J.V. Tucker and J.I. Zucker (2014):
   Computability of operators on continuous and
   discrete time streams. Computability, 3:9-44.

REFERENCE \cite{TZ14}

   This is actually a sequel to [289].
   Briefly: in [289] we proved that (under certain reasonable conditions)
   the fixed point function derived from a network is *continuous*.
   This is important in light of Hadamard's Principle, which, 
   in the present context, can be (re-)stated in the form:
   for a model of a physical system to be acceptable, the behaviour 
   of the model must depend continuously on the data.
   (See Discussion 4.2.14 in [289].)

   In the present paper, we prove (again, under reasonable conditions) 
   that the fixed point function is also *computable* 
   (according to some well-known  digital computation models).

   \textcolor{red}{AMAURY}
}


\AFAIRERELIREE{Tucker+Zucker+Costa+Beggs}

An approach to define
computations by analog  models in general  is to consider networks of analog modules
connected by channels \cite{TZ07}, processing data from a metric space $A$, and operating with
respect to a global continuous clock $\mathbb{T}$. The inputs and outputs of the network are
continuous streams ($C=\mathbb{T}\to A$) and the input-output behaviour of the network,
usually with external parameters, is modeled by a function $\phi:C^r\times A^p\to C^q$.
The authors focus on the important case where the modules are \emph{causal}, that
is the output at time $t$ only depends on the inputs over $[0,t]$.

Equational specifications of such circuits, as well as their
semantics, are given by fixed points of operators over the space of
continuous streams. Under suitable hypotheses, this operator is
contracting and an extension of the Banach fixed point theorem for
metric spaces guarantees existence and unicity of the fixed
point. Moreover, that fixed point can also be proven to be
continuous and \textit{concretely} computable whenever the basic
modules are.    

A general framework to deal with fixed point techniques in analog
systems, based on Fréchet spaces, has recently been developed in
\cite{poccas2016fixed}.

An abstract model of computability over
data stream using a high-level programming language, an extension of ``while'' programs,
over abstract data-types (multi-sorted algebras) has been considered in \cite{TZ14}. The authors analyze
when concrete and abstract models are equivalent.


\subsubsection{Formalization and Proof Theory for Cyberphysical Systems
}{\QUID{Olivier}}\label{sec:cps}

\AFAIRERELIREE{André Platzer}

\feedbackNIOK{André Platzer}{Hi Olivier and Amaury,

This sounds very interesting. I'm presently quite busy with finalizing this textbook \url{http://lfcps.org/lfcps/}

It looks like the discrete=continuous=hybrid proof theory with its constructions from LICS'12 \url{http://dx.doi.org/10.1109/LICS.2012.64} is also relevant for the relations on computations. Oh and in terms of verification this also gives you relative decidability.

For a hybrid distributed computation model you may also be interested in LMCS'12 \url{http://dx.doi.org/10.2168/LMCS-8(4:17)2012}

Hope this helps. And please let me know when the book appears.

Best wishes,
Andre

\textcolor{red}{OLIVIER}
}

A whole theory for the analysis, logic and proofs for cyberphysical
systems has been developed \cite{Platzer18}. Rich logical theories
to cover both continuous differential equation dynamics and discrete
changes involved in cyberphysical or hybrid systems have been proposed
and proven to be sound and complete
\cite{platzer,platzer2012complete,platzer2015uniform, platzer2012complete}. These logical
foundations, built incrementally in a series of articles, are now
forming an elegant proof-theoretical bridge aligning the theory of
continuous systems with the theory of discrete systems \cite{platzer2010logical, platzer2016logic}. They are the
basis of theorem provers \emph{KeYmaera} and \emph{KeYmaera X}, which demonstrated
their impact on various concrete applications. For an overview of all the existing
works along these lines of research at both theoretical and
practical levels, refer to \cite{platzer2016logic,Platzer18}.

\subsection{$\R$-recursion Theory}{\QUID{Olivier}}
\label{sec:rrecursive}

\AFAIRERELIREE{Mycka  + Costa + Campagnolo}
$\R$-recursive functions were introduced as a theory
of recursive functions on the reals built in analogy with classical
recursion theory to deal with conceptual analog computers operating in
continuous time \cite{Moo95b}. Moore considered the smallest class of functions
obtained from constants $0$ and $1$, projections and closed by
composition, integration and minimization.  He demonstrated that
various non-recursively enumerable sets fall into the proposed hierarchy,
defined according to the number of minimizations involved.

The initial theory \cite{Moo95b} suffers from some
mathematical difficulties when considering several nested
minimization operators, but its foundational ideas gave birth to
several lines of research, both at the computability and complexity
level.

At the computability level, since minimization is the operation that gives rise to
uncomputable functions, and difficulties, a natural question is to ask
if it can be replaced by some other natural and better defined operator
of mathematical analysis. This can be done by replacing
minimization by some limit operation \cite{MC04}.  In
addition, it is shown that the obtained hierarchy does not collapse \cite{LCM07,Loff07},
which implies that infinite limits and first
order integration are not interchangeable operations
\cite{CLM07}. A presentation and overview of the obtained real
recursive functions theory is presented  in
\cite{costa2009foundation}. 

The algebra of functions built without the minimization operator
only contains analytic functions and is not closed under iteration
\cite{CMC01}. Closure by bounded products,
bounded sums and bounded recursion has also been investigated.

However, if an arbitrarily smooth extension
$\theta$ to the reals of the Heaviside function is included in the set
of basic functions, then extensions to the reals of all primitive
recursive functions are obtained.  
More generally, several authors studied extensions of the algebra of
functions by considering an integration operator restricted to {\em
  linear} differential equations, and considering that such an
arbitrarily smooth extension $\theta$ was among the basic functions.
The obtained class contains extensions to the reals of all the
elementary functions \cite{Campagnolo,cam:moo:fgc:00,cam:umc:02,cam:tcs:04}. By adding a suitable basic
exponential iterate, extensions to the reals of all the functions of the classes
of the Grzegorczyk's hierarchy can also be obtained.

Instead of asking which computable functions over $\N$ have extensions
to $\R$ in a given function algebra, one can consider
classes of functions over $\R$ computable according to recursive
analysis, and characterize them precisely with function algebras.
This was done for elementarily computable functions \cite{BH05},
considering a restricted limit schema, and for computable functions
\cite{BH06}, considering a restricted
minimization schema.  

A generic framework for presenting previous results, based on the
notion of approximation, has been developed in \cite{CO08}.

At the complexity level, characterizations of complexity classes such as $P$ and $NP$
using $\R$-recursive algebra have been obtained \cite{MC06}.
This is done by ensuring that at every step of the construction of a
function in previous classes, each component cannot grow faster than a
quasi-polynomial. This allows to transfer classical questions from
complexity theory to the context of real and complex
analysis \cite{LCM07b, MC06,MC05Eatcs}.  



\subsection{Analog Automata Theories} 
\label{sec:continuousautomata}

\feedbackOK{Emmanuel Jeandel}{ 
(a) Vous pouvez (lire et) citer \cite{bozapalidis2003extending} 
Bozapalidis, Extending Stochastic and
Quantum Functions dans 5.3
\textcolor{red}{OLIVIER}
}

Several lines of research have been devoted to adapt classical
discrete automata theory to analog domains or continuous time.

Rabinovich and Trakhtenbrot have developed a \emph{continuous time automata}
theory in \cite{RabTra97,Trakhtenbrot99,Rabinovich03b}. In this approach, automata are not considered as
language recognizers, but as computing operators on signals, that is
on functions from the non-negative real numbers to a finite alphabet,
seen as channel's states. For a presentation as well as extensions, see \cite{Franc02}.

Another approach has been considered, where ODEs equipped with a
tape-like function memory are used to recognize sets of piecewise
continuous functions, yielding a Chomsky-like hierarchy \cite{Ruo04}.
\TODO{Il y a moore aussi.}

A class of recognizable functions extending stochastic and quantum
functions, and with a cutpoint theorem similar to the one for
probabilistic automata theory and with nice closure properties has
been determined in \cite{bozapalidis2003extending}.



{\QUID{Amaury (premier jet d'Olivier)}}

\emph{Topological automata} were introduced
as a generalization of previous definitions of probabilistic and
quantum automata \cite{jeandel2007topological}: deterministic or nondeterministic probabilistic and
quantum automata are proven to recognize only regular languages with
an isolated threshold. A topological theory
of continuous-time automata has also been developed \cite{messick2016compactness}: the basic idea is to replace finiteness
assumptions in the classical theory of finite automata by compactness
assumptions. Some existence results and a Myhill-Nerode theorem is
obtained in this framework covering both finite automata and
continuous automata.

{\QUID{Amaury (premier jet d'Olivier)}}

\label{sec:automatameer}
\emph{Generalized finite automata} working over arbitrary structures have
been introduced \cite{gandhi2012finite}. Structural properties of
accepted sets as well as computational hardness of classical questions for
this model have been investigated \cite{meer2015generalized}.  A
restricted version of the model has also been considered
\cite{meer2016periodic}: it
is demonstrated  to satisfy a pumping lemma, and to yield decidability
results closer to what would be expected for a notion of finite
automata over arbitrary structures, such as the reals. 

All previous approaches are not easily connected, and are rather
independent views and models.



\section{Analysing the Power and Limitations of Analog Computations}\label{sec:power_and_limits_of_models}

\subsection{Neural Networks }{\QUID{Olivier}}

\feedbackOK{Wolfgang Maass}{ Dear Olivier:
JE FAIS l'IMPASSE CAR JE VOIS PAS OU METTRE CA.
It would seem to me that \cite{MJS07}
provides additional info on analog computation in neural networks, and
\textcolor{red}{OLIVIER A BAFFER}
}

\label{sec:nn}

\AFAIRERELIREE{Jérémie Cabessa} The computational power of artificial
neural networks has been investigated intensively in the 90's in many
papers: see the instructive survey \cite{SOSurvey} and more recently
\cite{HDRCabessa}.  Finite analog recurrent networks with integer
weights are known to be essentially equivalent to finite automata,
while finite analog recurrent networks with rational weights can
simulate arbitrary Turing machines step by step \cite{SS95}. When
considering arbitrary real weigths (possibly non-computable) one
obtains the non-uniform versions of the associated complexity and
computability classes. In particular, one gets models that can decide
predicates or sets that are not computable by Turing machines
\cite{SS94}. For a full discussion, see \cite{Sie99}.
Improvements on the mapping from a Turing machine to a recurrent
artificial neural network have recently been proposed
\cite{carmantini2015turing}. It has also been proven that the
question $P=? NP$ relativizes naturally to similar questions
related to artificial neural networks with real weights
\cite{DBLP:journals/tcs/CostaL13}.

\feedbackOK{Jérémie}{[...] Tu peux citer les papiers journaux plutôt que les conf, c'est plus complet:
IJNS \cite{cabessa2014super}, JCSS \cite{cabessa2016expressive}, PLoS One\cite{cabessa2014attractor} \textcolor{red}{OLIVIER}}
Previous results have been extended to the case of evolving recurrent
neural networks, i.e., networks where the synaptic weights can be updated
at each discrete time step
\cite{cabessa2011evolving,cabessa2014super}. Interactive recurrent
neural networks have been considered in
\cite{cabessa2012computational,cabessa2013super}. 

When subjected to some infinite binary input stream, the Boolean
output cells necessarily exhibit some attractor dynamics.  A
characterization of the expressive powers of several models of
Boolean, sigmoidal deterministic, and sigmoidal nondeterministic
first-order recurrent neural networks, in relation with their
attractor dynamics has been established in 
\cite{cabessa2015computational, cabessa2014attractor, cabessa2016expressive}. This extends results about their expressive power 
characterized in terms of topological classes in the Borel hierarchy
over the Cantor space \cite{cabessa2012expressive}. 

The computational power of spiking neuron models has been investigated
in a series of papers: refer to \cite{SOSurvey,MB98} or \cite{ghosh2009spiking} for surveys.

\subsection{Physical Oracles }{\QUID{Amaury}}
\label{sec:physical}

In a recent series of papers, started by \cite{beggs2008computational}, Beggs, Costa, Po\c{c}as and Tucker
consider various dynamical systems doing computations involving discrete data and physical
experiments. Such systems can be modeled
by Turing machines with oracles that correspond to physical processes. A good summary of
their results can be found in \cite{ambaram2017analogue}. They analyze the
computational power of these machines and more specifically the limits of polynomial
time computations. They show that for a broad class of physical oracles, an upper bound
on polynomial time computations is the non-uniform complexity class\footnote{Let $\mathcal{B}$
is a class of sets and $\mathcal{F}$ a class of functions. The advice class $\mathcal{B}/\mathcal{F}$
is the class of sets $A$ such that there exists $B\in\mathcal{B}$ and $f\in\mathcal{F}$
such that for every word $w$, $w\in A$ if and only if $\langle w,f(|w|)\rangle\in B$.
We use $log$ to denote the class of functions $f$ such that
$|f(n)|=\mathcal{O}(\log n)$ as $n\to\infty$.
We use $poly$ to denote the class of functions $f$ such that
$|f(n)|=\mathcal{O}(n^k)$ for some $k$ as $n\to\infty$.
Thus $P/poly$ corresponds to a non-uniform polynomial size advise. The class $BPP//\mathcal{F}\star$ is
the class of sets $A$ such that there is a probabilistic polynomial time Turing machine $\mathcal{M}$,
a prefix function $f\in\mathcal{F}\star$ (\emph{i.e.} $f(n)$ is a prefix of $f(n+1)$) and a constant $\gamma<\tfrac{1}{2}$ such 
that for every word $w$, on input $\langle w,f(|w|)\rangle$, $\mathcal{M}$ rejects
with probability at most $\gamma$ if $w\in A$, and accepts with probability at most $\gamma$ if $w\notin A$.
See \cite{beggs2008computational} for more details on those classes.}
BPP//log$\star$, and that P/poly can be attained by using non-computable
analog-digital interface protocols.

Without giving too much details, in this model the Turing machine has access to
a physical oracle, such as \cite{beggs2007experimental}. Whenever the machine wants to call the oracle, it writes some input
on the oracle tape and enters a special oracle state. The machine is then suspended
and a physical experiment starts. Importantly, to each machine is associated a
time schedule function $T$, usually a time-constructible function. If the experiments
finishes in time less than $T(|z|)$, where $z$ is the content of the oracle tape, the machine resumes in one of
finitely many states encoding the outcome of the experiment (typically \emph{YES} or \emph{NO}).
Otherwise, the experiment is stopped and the machine resumes in a special state
\emph{TIMEOUT}.

The authors analyzed many kinds of physical oracles and identified
three major forms of experiments. In most cases, the experiment encodes a single
real number $a$ and the oracle tape encodes a rational $x$. Each class of experiments
specifies not only the behavior of the experiments but also how long it takes to
complete.
\begin{itemize}
\item\textbf{Two-Sided Case:} the experiment takes time $t(x,a)=C/|x-a|^d$ for $x\neq a$ and
    answers YES if $x<a$ and NO if $x>a$. It always times out if $a=x$. A typical
    example is a scale, where the time needed to detect movement to either side
    is inversely exponential to the difference between the masses.
\item\textbf{Threshold Case:} the experiment takes time $t(x,a)=C/|x-a|^d$ for $x>a$
    and always answers YES. It always times out if $x\leqslant a$. Several examples
    including the photoelectric effect and Rutherford scattering are given by the
    authors.
\item\textbf{Vanishing Case:} the experiment can only detect if $x
  \neq a$
    and times out if $x= a$. However, given two experiments $x,x'$, we can detect
    which of $x$ and $x'$ is closer to $a$. Informally, the reader can think of an experiment
    where we can measure $\alpha (x-a)^2$ (with $\alpha$ some unknown parameter): we
    cannot detect the sign of $x-a$ but for two experiments $x,x'$, $\alpha$ are the same so
    we can compare $\alpha (x-a)^2$ to $\alpha (x'-a)^2$. Examples include some modified Wheatstone bridge
    and Brewster's angle experiment.
\end{itemize}
Orthogonally to the type of physical experiment, the authors also study variants
on how precise the experiment is with respect to $x$. In the \emph{infinite} precision
case, the experiment is done exactly with $x$. In the \emph{unbounded} precision case,
the experiment is done with $x'\in[x-\varepsilon,x+\varepsilon]$ (drawn uniformly at
random and experiments are independent) where $\epsilon$ is part of the input and can be arbitrarily small (but not zero).
In the \emph{fixed} precision case, $\epsilon$ is a constant of the experiment and
cannot be changed.

More recently, the authors have developed a hierarchy for BPP//log$\star$
based on counting calls to a non-deterministic physical oracle modeled by a random
walk on a line \cite{Beggs2017}.

A general framework for describing physical computations, covering
above approaches has been proposed in \cite{whyman2016physical}.

\subsection{On the Effect of Noise on Computations}{\QUID{Olivier}}

\feedbackOK{Wolfgang Maass}{ Dear Olivier:
It would seem to me that \cite{MS99}
provides additional info on the section about effects of noise.

\textcolor{red}{OLIVIER A SEMI-BAFFER}
}

\AFAIRERELIREE{Braverman, Rojas}
\AFAIRERELIRE{ Schneider?}
The invariant (ergodic) measure of dynamical systems has been proven
to remain computable even if a small amount of
noise is introduced in the system \cite{braverman2012noise}. This
demonstrates that while some dynamical systems have been proven to
simulate Turing machines, the presence of noise forbids
uncomputability. Actually, when considering a compact domain, it has been proven
that analog neural nets with gaussian or other common noise
distributions cannot even recognize arbitrary regular languages
\cite{MO97,MS99}. 

Some tight bounds on the space-complexity of computing the ergodic
measure of a low-dimensional discrete-time dynamical systems affected
by a Gaussian noise have been obtained \cite{braverman2015tight}.

Previous results have implied some limitations on the ability of physical systems to
perform computations, namely on their ability to store information.
Memory is roughly defined as the maximal amount of information that
the evolving system can carry from one instant to the next, and is
demonstrated not to be able to grow too fast. 

This is discussed to be in favor of the following postulate for
physical computations. If a physical system has memory $s$ available
to it, then it is only capable of performing computations in the
complexity class $SPACE(poly(s))$ even when provided with unlimited
time \cite{braverman2015space}. This has been proven to hold for several
classes of dynamics and noise models over a compact domain.\TODO{A mon
  avis le compact est important (et un peu tricher)}




\subsection{Complexity theories for Analog Computations}

\AFAIRERELIREE{Marco Gori et Klaus Meer}
There have been several attempts to build a complexity theory for
continuous time systems (but not intended to cover generic ODEs). 

This includes the theory where the global minimizers of particular
energy functions are supposed to give solutions of some problem \cite{gori2002step, gori2007some}. The
structure of such energy functions leads to the introduction of
problem classes $U$ and $NU$, with the existence of complete problems
for theses classes.

Another attempt is focused on a very specific
type of systems: \emph{dissipative flow models} \cite{HSF02}. This theory has been used in
several papers from the same authors to study a particular class of
flow dynamics \cite{BFFS03} for solving linear programming problems.

Neither of the previous two approaches is intended to cover generic ODEs, and
none of them is able to relate the obtained classes to classical
classes of computational complexity, unlike the approach presented
in Section \ref{sec:pode}.

\subsection{Chemical Reaction Networks }{\QUID{Olivier}}

\label{sec:SCRN}

\AFAIRERELIREE{Dave Doty + Dave Soloveichich + Cook }

Recent years have seen an important literature about the computational
power of \emph{chemical reaction networks} (CRN)
\cite{soloveichik2008computation, cook2009programmability}. The model
is built by analogy with concrete chemistry, see also
\cite{reus2016limitschapter} for a presentation.  A program
corresponds to a set of chemical rules over a finite set of species,
abstracting from the matter conservation laws, considering well-mixed
solutions, and ignoring spatial properties of the molecules. They
can be seen as natural extensions of population protocols
discussed in Section \ref{sec:populationprotocol}: here reactions are
not assumed to be between pairs of agents/molecules, and reaction
rates can possibly be considered.

  One can distinguish various dichotomies of computations in CRNs \cite{cummings2014probability,SlidesSoloveichik}:
  the number of molecules can either be considered as \emph{discrete} or
  \emph{continuous}. Computations can be considered either \emph{uniform} (the same
  CRN handles all inputs) or \emph{non-uniform} (for each input size, a different CRN is
  considered). In the \emph{deterministic} setting the output is considered
  to be guaranteed, while in the \emph{probabilistic} setting some
  probability of error is considered.  The \emph{halting approach} considers
  notions of acceptance where agents irreversibly produce an answer,
  while in the \emph{stabilizing approach} the population eventually
  stabilizes to an answer, in the spirit of the acceptance criteria for
  population protocols.

  CRNs have clear connexions to various other models, including
  Petri nets, and vector addition systems
  \cite{cook2009programmability}.  These models can be classified as
  \emph{unordered} with respect to classical models of computability such as
  Turing machines where an order on instructions is considered. The
  presence of reaction rates or probabilities is a way to provide back
  some order on instructions, while forgetting order in classical
  models of computability gives rise to similar models
  \cite{cook2009programmability}.

 This abstract model is also motivated by concrete implementations
 using an experimental technique called DNA strand displacement
 \cite{soloveichik2010dna,cardelli2009strand,QSW11dna,chen2013programmable}. Implementations
 with proteins are also considered \cite{CMSB17}. \TODO{Meillere
   ref sur protéines.}
\AFAIRERELIREE{Leroux}
\AFAIRERELIRE{Esparza? coauteurs?.}

  The computational power of CRNs over many of these
  variants has been studied in several papers
  \cite{soloveichik2008computation,chen2014deterministic,cummings2014probability}. 
  It has been proven, using in particular
  the strong technical result from \cite{esparza2016verification}, that chemical reaction
  networks turn out to be a robust model from a computability point
  of view \cite{brijder2016robustness} in the sense that variations on the notion of output
  conventions for error-free computations basically yields either 
  Turing machines or a model whose power is equivalent to
  population protocols discussed in Section
  \ref{sec:populationprotocol}, that is to say whose computational
  power is limited to semi-linear sets.

Lower
  bounds on the time required to build rules that would act as a
  ``timer'' have been obtained \cite{doty2014timing}, answering the
  open question on the ability of population protocols to perform a fast
  leader election in the probabilistic settings, as well as their
  ability to simulate arbitrary algorithms from a uniform initial
  state.

  The case of the continuous settings with the usual natural semantics
  has been an open problem before being very recently solved \cite{CMSB17} using the
  theory discussed in Section \ref{sec:pode}. The
  computational power of a variant where reaction rates are abstracted
  was previously characterized \cite{chen2014rate}.






\section{Computations by Polynomial Ordinary Differential Equations} 
\label{sec:pode}

Polynomial ordinary differential equations have been shown to be
closely related to the GPAC of Shannon discussed in Section
\ref{sec:gpac} and to have a very rich theory. In particular, they may
now be considered for continuous time models as the equivalent of
Turing machines for discrete time models in many aspects.

\subsection{GPAC and Polynomial Ordinary Differential Equations }{\QUID{Amaury}}\label{sec:theory_generable_functions}

The GPAC, presented in Section \ref{sec:gpac}, can also be presented
in terms of polynomial ordinary differential equations.  
More precisely, $y$ is \emph{generable by a GPAC} if and only
if there exist functions $z=(z_1, \ldots, z_n)$, a vector
of polynomials $p$ and an initial condition $z_0$ such that
\[y\equiv z_1,\qquad z(0)=z_0,\quad\text{and}\quad z'=p(z).\]
In other words, $y$ is a component of the solution of a system of polynomial differential equations.
One advantage of this presentation is that it eliminates all the problems related to
the domain of definition and the solution is necessarily unique. In particular,
the solutions are always analytic functions.

This fact actually has a complicated history because of gaps in proof
found by various authors and the refinements in the definition of the
GPAC:  Shannon, while introducing his model \cite{Sha41}, proved that any function $y$ generated
by a GPAC is solution of a polynomial Differential Algebraic Equation
(pDAE), that is there exists a polynomial $p$ such that
\begin{equation}\label{eq:gpac_poly_dae}
p(y,y',\ldots,y^{(n)})=0
\end{equation}
for some integer $n$. The converse inclusion was claimed by Shannon,
which led some authors to define the GPAC in terms of differential
equations directly \cite{PER74}.  However this equivalence requires
some care because of the domain of definition, and furthermore, DAEs do not
necessarily have a unique solution, which complicates this
approach. It was later realized that unary functions
generated by a GPAC (with some restrictions to fix earlier problems)
are exactly components of polynomial Initial Value Problems (pIVP)  \cite{GC03}.
Several variations on the GPAC circuits were explored and proven to be
all equivalent, which essentially shows that the above notion is
robust and probably the right definition to be considered \cite{Gra04}.

A famous statement, due to 
Rubel \cite{Rub81}, shows that polynomial Differentially
Algebraic Equations (pDAEs) 
are \emph{universal} for continuous functions. More precisely,
there is a universal pDAE of order 4 whose
solutions can approximate any continuous function with arbitrary
precision \cite{Rub81}. This work has recently been improved
\cite{couturier2016construction} to hold for order  3. It has been
established very recently that this also holds for polynomial ordinary
differential equations  
\cite{ICALP2017}.

\subsection{GPAC Generable Functions}

The class of generable functions is particularly robust because if $f$ and $g$
are generable then $f\pm g$, $fg$, $\tfrac{f}{g}$ and $f\circ g$ are also generable.
Moreover, if $f$ is generable and $y$ satisfies the differential equation
$y'=f(y)$ then $y$ is also generable. Shannon's original work was intended for circuits
with several inputs, but this was never formalized. Recent work 
provides a proper description of  the class of generable functions over multi-dimensional domains and shows
that it also enjoys many closure properties \cite{2016arXiv160200546B}.

\subsection{GPAC computability}{\QUID{Amaury}}\label{sec:polynomial_odes}

The fact that the functions generated by the GPAC (i.e. generable functions) are analytic (or
$C^\infty$ for the more relaxed models, without external inputs) has
historically been seen as a limitation of the GPAC. 




This has been proven to be an artefact of the model
and an alternative notion of computability based on
polynomials ODEs has been proposed  \cite{BCGH07}. In fact the authors prove that the GPAC and Turing machines
are equivalent. The fundamental idea is that while generable functions
correspond to ``\emph{real-time computability}'' (\emph{i.e.}, the system
computes the answer instantaneously), considering a notion similar to
the one used for Turing machines (\emph{i.e.}, the system evolves and
``converges effectively'' to
the answer) yields a computational power equivalent to that of
Turing machines. 

This equivalence between the GPAC and Computable Analysis
can be
reformulated as follows:

\begin{theorem}[Equivalence between GPAC and Computable Analysis, \cite{BCGH07}]
A function $f:[a,b]\to\R$ is computable (in the sense of Computable Analysis) if
and only if there exists an integer $d$ and a vector of polynomials $p:\R^d\to\R^d$
with rational coefficients such that for any $x\in[a,b]$, the unique solution $y:\Rp\to\R^d$ to
\[y(0)=(0,\ldots,0,x),\qquad y'(t)=p(y(t))\]
satisfies that
\[|f(x)-y_1(t)|\leqslant y_2(t)\quad\text{and}\quad y_2(t)\to 0\text{ as }t\to\infty.\]
\end{theorem}
In other words, a function $f$ is computable if there is a GPAC that, on input $x$,
has one of its components ($y_1$) converge to $f(x)$. Not only it converges, but
another component ($y_2$) gives a bound on the error between $f(x)$ and $y_1$.

This unexpected connection between Turing computability and the GPAC was recently
refined \cite{ICALP2016,JournalACM2017} at the level of complexity, with a characterization of the class P and
polynomial time computable real functions. 

A difficult point in the
context of analog computability, and in the GPAC in particular is to define
a notion of complexity that makes sense and is sufficiently robust:
see discussions in Section \ref{sec:zeno}. 
In fact, the intuitive notion of
measuring the complexity based on the convergence rate (\emph{i.e.} how fast $y_1(t)\to f(x)$)
does not work. 

One recently proposed solution to this
problem is to measure the complexity using the \emph{length of the curve $y$} instead
of time. This process is illustrated in Figure~\ref{fig:gpac_converge_length}.
We recall that the length of a curve $y\in C^1(I,\R^n)$ defined over
some interval $I=[a,b]$ is given by
$\glen{y}(a,b)=\int_I\twonorm{y'(t)}dt,$

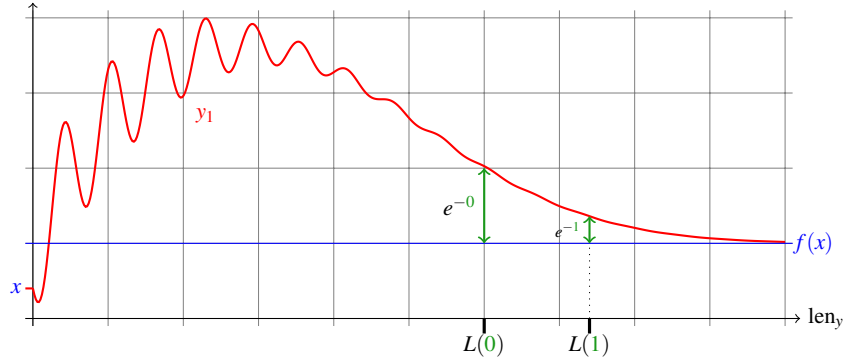
\begin{figure}
\centering
\begin{tikzpicture}[domain=1:11,samples=500,scale=1]
    \draw[very thin,color=gray] (0.9,-0.1) grid (11.1,4.1);
    \draw[->] (1,-0.1) -- (1,4.2);
    \draw[->] (0.9,0) -- (11.2,0) node[right] {$\glen{y}$};
    \draw[color=red,thick] plot[id=fn_1] function{
        1+exp((x-1)*(7-x)/10)+(1+sin(10*x))/(1+exp(x-3))-2*exp(-(x-1))
        };
    \draw[blue] (0.9,1) -- (11.1,1);
    \draw[right,blue] (11,1) node {$f(\textcolor{blue}{x})$};
    \draw[color=red,thick] (1.0,0.4) -- (0.9,0.4);
    \draw[red] (0.95,0.4) node[left] {$\textcolor{blue}{x}$};
    \draw[color=red] (3.3,2.7) node {$y_1$};
    \draw[very thick] (7.0,0) -- (7.0,-0.2);
    \draw[<->,darkgreen,thick] (7.0,1) -- (7,2) node[midway,black,left] {$e^{-\textcolor{darkgreen}{0}}$};
    \draw (7,-0.1) node[below] {$L(\textcolor{darkgreen}{0})$};
    \draw[<->,darkgreen,thick] (8.4,1) -- (8.4,1.36) node[midway,black,left] {$\scriptstyle e^{-\textcolor{darkgreen}{1}}$};
    \draw[dotted] (8.4,0) -- (8.4,1);
    \draw[very thick] (8.4,0) -- (8.4,-0.2);
    \draw (8.4,-0.1) node[below] {$L(\textcolor{darkgreen}{1})$};
\end{tikzpicture}
\caption{Graphical representation of analog computability (Theorem~\ref{th:analog_poly_length}): on input $x$, starting from initial condition $(x,0,\ldots,0)$,
the polynomial ordinary differential equation $y'=p(y)$ ensures that $y_1(t)$ gives $f(x)$ with accuracy better than $e^{-\mu}$
as soon as the length of $y$ (from $0$ to $t$) is greater than $L(\mu)$.
Note that we did not plot the other variables $y_2,\ldots,y_d$ and the horizontal
axis measures the length of $y$ (instead of the time $t$).\label{fig:gpac_converge_length}}
\end{figure}

\begin{theorem}[Equivalence between GPAC and CA (Complexity), \cite{JournalACM2017}]\label{th:analog_poly_length}
A function $f:[a,b]\to\R$ is computable in polynomial time (in the sense of Computable Analysis) if
and only if there exists a polynomial $L:\Rp\to\Rp$, an integer $d$ and a vector of polynomials $p:\R^d\to\R^d$
with coefficients in $\Q$,
such that for any $x\in[a,b]$, the unique solution $y:\Rp\to\R^d$ to
\[y(0)=(x,0,\ldots,0),\qquad y'(t)=p(y(t))\]
satisfies for all $t\in\Rp$:
\begin{itemize}
\item for any $\mu\in\Rp$, if $\glen{y}(0,t)\geqslant L(\mu)$ then $|f(x)-y_1(t)|\leqslant e^{-\mu}$,
\item $\infnorm{y'(t)}\geqslant 1$.
\end{itemize}
\end{theorem}
In other words, the precision of $y_1$ increases with the length of the curve. More precisely,
as soon as the length between $0$ and $t$ is at least $L(\mu)$, the precision is at least
$e^{-\mu}$. Notice how rescaling the curve would not help here since it does not
change the length of $y$. The second condition on the derivative of $y$ prevents some
pathological cases and ensures that
curve has infinite length, and thus that $y_1$ indeed converges to $f(x)$. It is
possible to extend this equivalence to multivariate functions and unbounded input domains
such as $\R$, by making $L$ take into account the norm of $x$.

It also possible to define the class $P$ directly in terms of differential equations,
by encoding words with rational numbers. Again the length plays a crucial role,
but since a differential equation does not ``stop'', the component
$y_1$ is used to signal that it accepts ($y_1\geqslant 1$) or rejects ($y_1\leqslant -1$).
Figure~\ref{fig:analog_recognizability} illustrates this process.

\begin{theorem}[Analog characterization of P, \cite{JournalACM2017}]\label{th:analog_p}
A language $\mathcal{L}\subseteq\{0,1\}^*$ belongs to P, the class of polynomial
time decidable languages, if and only if there exist 
a polynomial $L:\N\to\N$, an integer $d$ and a vector of polynomials $p:\R^d\to\R^d$
with coefficients in $\Q$, such that for all words $w\in\{0,1\}^*$,
the unique solution $y:\Rp\to\R^d$ to
\[y(0)=(0,\ldots,0,|w|,\psi(w)),\qquad y'(t)=p(y(t))\]
where $\psi(w)=\sum_{i=1}^{|w|}w_i2^{-i}$, satisfies for all $t\in\Rp$:
\begin{itemize}
\item if $|y_1(t)|\geqslant1$ then 
$|y_1(u)|\geqslant1$ for all $u\geqslant t\geqslant 0$ (and similarly for $|y_1(t)|\leqslant-1$),
\item if $w\in\mathcal{L}$ (resp. $\notin\mathcal{L}$) and $\glen{y}(0,t)\geqslant L(|w|)$
then $y_1(t)\geqslant1$ (resp. $\leqslant-1$),
\item $\infnorm{y'(t)}\geqslant 1$.
\end{itemize}
\end{theorem}

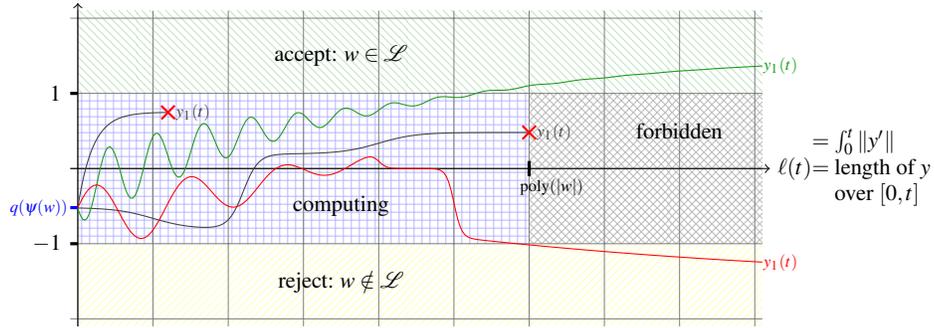
\begin{figure}
\begin{center}
\begin{tikzpicture}[domain=1:10.1,samples=300,scale=1]
\fill[pattern=north west lines, pattern color=darkgreen!25!white
] (1,1) rectangle (10.1,2.1);
\fill[pattern= north east lines, pattern color=yellow!25!white
] (1,-1) rectangle (10.1,-2.1);
\fill[pattern= grid,pattern color =blue!25!white
] (1,-1) rectangle (7,1);
\fill[pattern= crosshatch, pattern color=black!25!white
] (7,-1) rectangle (10.1,1);
\draw[very thin,color=gray] (0.9,-2.1) grid (10.1,2.1);
\draw[->] (1,-2.1) -- (1,2.2);
\draw[->] (0.9,0) -- (10.2,0) node[right] {
    $\displaystyle\begin{array}{@{}r@{}l@{}}
        &=\int_0^t\infnorm{y'}\\
        \ell(t)&=\text{length of }y\\
        &\hspace{1em}\text{over }[0,t]\end{array}$};
\draw[very thick] (1,1) -- (0.9,1) node[left] {$1$};
\draw[very thick] (1,-1) -- (0.9,-1) node[left] {$-1$};
\draw[very thick] (7,0.1) -- (7,-0.1) node[below=-.3em,xshift=1em] {$\scriptstyle\poly(|w|)$};
\draw[thick
] (4.5,1.5) node {accept: $w\in\mathcal{L}$};
\draw[thick
] (4.5,-1.5) node {reject: $w\notin\mathcal{L}$};
\draw[thick
] (4.5,-.5) node {computing};
\draw[thick
] (9,.5) node {forbidden};
\draw[color=blue,very thick] (1,-.52) -- (0.9,-.52) node[left=-.3em] {$\scriptstyle q(\psi(w))$};
\draw[color=black!75!white,domain=1:7,
        postaction={decorate,decoration={markings,mark=at position 1 with {
            \draw[red,thick] (-.3em,-.3em) -- (.3em,.3em); \draw[red,thick] (-.3em,.3em) -- (.3em,-.3em);}}}]
        plot[id=ar_timeout] function{-0.52-(1+tanh((x-2)*2))/7+(1+tanh((x-3.2)*5))/2+(1+tanh((x-5)*2))/7} node[right] {$\scriptstyle y_1(t)$};
\draw[color=black!75!white,domain=1:2.2,
        postaction={decorate,decoration={markings,mark=at position 1 with {
            \draw[red,thick] (-.3em,-.3em) -- (.3em,.3em); \draw[red,thick] (-.3em,.3em) -- (.3em,-.3em);}}}]
        plot[id=ar_stop] function{.75-(0.75+0.52)*exp((1-x)*5)} node[right] {$\scriptstyle y_1(t)$};
\draw[color=darkgreen] plot[id=ar_accept] function{(x/3+sin(10*x))/(1+exp(x-3))/2+3./2.*tanh(x/5-1./2.)} node[right=-.3em] {$\scriptstyle y_1(t)$};
\draw[color=red] plot[id=ar_reject] function{((-1.1-x/10-sin(5*x-1.5))/(1+exp(x-3))/2+1./2.*tanh(x/5-1./2.))*(1+tanh((5-x)*10.))/2.
    -(1-tanh((6-x)*10))/2.*3./2.*tanh(x/8.5)} node[right=-.3em] {$\scriptstyle y_1(t)$};
\end{tikzpicture}
\end{center}
\caption{Graphical representation of the analog characterization of P (Theorem~\ref{th:analog_p}).
The green trajectory represents an accepting computation, the red a rejecting one, and the gray are invalid
computations. An invalid computation is a trajectory that is too slow (thus violating the technical condition),
or that does not accept/reject in polynomial length. Note that we only represent the first
component of the solution, the other components can have arbitrary behaviors.}
\label{fig:analog_recognizability}
\end{figure}

One clear interest of the previous statements is that they provide a way
to define classical concepts from the theory of computation (computable
function, polynomial time computable functions) only using concepts
from analysis, namely polynomial ordinary differential equations. 









\section{Acknowledgements}

\olivierOK{Personnes contactées:
  \begin{itemize}
  \item Bernd Ulmann
\item MacLenann
\item Ulf Hashagen,
\item Daniel Graça
\item Felix Costa
\item Hava Siegelmann
\item Orponen
\item Jérémie Cabessa
\item Goran Frehse
\item Laurent Doyen
\item Klaus Meer
\item ? Gassner
\item Patricia Bouyer
\item Bernard Chazelle
\item Pietro Milici
\item ? Copeland
\item Maël Pegny
\item Manuel Campagnolo
\item Eugène Asarin
\item Florent Franchette
\item Gilles Dowek
\item Cris Moore
\item ? Tucker
\item Julien Cerevelle
\item Jack Lutz
\item Edwing Beggs
\item Nachum Dershowitz
\item Jérôme Durand Lose
\item ? Zucker
\item François Fages
\item Vincent Blondel
\item Emmanuel Hainry
\item Emmanuel Jeandel
\item Mark Braverman(n?)
\item Cristobal Rojas
\item André Platzer
\item Jerzy Mycka
\item Dave Doty
\item David Soloveichik
\item Matthew Cook
\item Erik Winfree
\item Jérôme Leroux
\item Marco Gori
\item Wolfgang Maass
\item Keijo Ruohonen
\item Laurent Fribourg
\item James Worell
\item Joel Waknine?
\item Marie-Jo Durand-Richard
\item Jonathan Mills
\item Andrew Adamatzky
\item Javier Esparza
\item Jean-Charles Delvenne (via Vincent Blondel)
\item Sylvie (via Eric Goubault)
\item Jack Lutz
\item Cameron Beebe
  \end{itemize}
}

\olivierOK{Ont promis une réponse, mais pas venue:
  \begin{itemize}
  \item Patricia Bouyer
\item Julien Cervelle
\item Dave Doty
\item Emmanuel Hainry
\item Bernard Chazelle
\item Christine Gassner
\item Jean-Charles Delvenne
  \end{itemize}
}

\olivierOK{Réponses avec encouragements:
(plus (liste non disjointe) ceux cités plus bas dans remerciements)
  \begin{itemize}
  \item Keijo Ruohonen
  \item John Tucker
  \item Jeffery Zucker
\item Amir Ali Amhadi 
  \end{itemize}
}

\olivierOK{Apparu entre temps (mais on en parle pas)
  \begin{itemize}
  \item 
\url{
http://scholar.google.com.br/scholar_url?url=https://arxiv.org/pdf/1610.09435&hl=fr&sa=X&scisig=AAGBfm13Qu9h76Al7zp-zIGf4UuHnHrSWQ&nossl=1&oi=scholaralrt}
\end{itemize}

}

}

\feedbackOK{Amaury}{
Amaury:Je ne l'ai pas lu en détail mais il donne un argument, basé sur
le problème à trois corps, de pourquoi on ne peut pas résoudre des ODE
en temps polynomial (avec la définition habitelle, ie pas la
longueur).
\url{
http://pca.pdmi.ras.ru/2016/abstracts_files/complexity3bp.pdf}
}

We would like to thank several people for various feedbacks, and many helpfull comments: 
Andrew Adamatzky,
Amir Ali Amhadi,
Cameron Beebe,
J\'er\'emie Cabessa,
Liesbeth Demol,
Dave Doty,
Laurent Doyen,
J\'er\^ome Durand-Lose,
Marie-Jo Durand-Richard,
Javier Esparza,
François Fages,
Goran Frehse,
Laurent Fribourg,
Daniel Gra\c{c}a,
Eric Goubault, 
Emmanuel Hainry,
Emmanuel Jeandel, 
Manuel Lameiras Campagnolo,
J\'er\^ome Leroux,
Jack Lutz,
Wolfgang Maass,
Klaus Meer,
Pekka Orponen,
Andr\'e Platzer,
Ma\"el P\'egny,
Crist\'obal Rojas, 
Keijo Ruohonen,
Zoltan Toroczkai,
John Tucker,
Bernd Ulmann,
Erik Winfree, 
Jeffery Zucker,
as well as two anonymous referees. 

Special thanks to Ulf Hashagen for many references related to history
of computing, Andrew Adamatzky and Keijo Ruohonen for several references and
comments, sometimes repeated literrally in this document, about some physical or
unconventional computing
models. 
        
Many thanks to J\'er\'emie Cabessa, J\'er\^ome Durand-Lose and Klaus Meer for many feedbacks
including typographic issues about a preliminary version of this
document.

\feedbackNIOK{Pekka Orponen}
{I suppose my personal main contributions in this research direction are the following:

1. Polynomial-size continuous-time Hopfield nets can simulate PSPACE
Turing machines (2003),
"https://doi.org/10.1162/089976603321192130". (This relates somewhat
to the discussion about solving SAT with dynamical systems in Sec 2.7:
the result shows that in fact even ODEs with Lyapunov-function
controlled dynamics can do much more.)

Apparemment: \cite{SO2003}

2. Finite-dimensional and bounded discrete-time dynamical systems with
noise have at most the power of finite automata (1997/98),
"https://doi.org/10.1162/089976698300017359". (Relates to Sec 6.3 on
the effect of noise on computations. The result was later extended by
Maass and Sontag to show that for e.g. Gaussian noise not even all
regular languages can be recognised.)

Apparemment \cite{MO97}

3. Finite 2D coupled map lattices are computationally universal
(Proceedings, 4th Workshop on Physics and Computation (Boston MA,
22.-24.11.1996),
"\url{http://interjournal.org/manuscript_abstract.php?2067910855}" and
"\url{http://citeseerx.ist.psu.edu/viewdoc/summary?doi=10.1.1.52.9278}". (Relates
to Sec 2.10 on spatial models. Somewhat interesting because this is a
finite planar nearest-neighbour model with very simple local
dynamics. Also because the model has been studied a lot by physicists
interested in PDEs and spatiotemporal chaos.)

C'est \cite{orponen1996universal} et

Best regards,
  Pekka

PS. The survey contains still contains quite a large number of typos.

}

\feedbackOK{Amaury}{
\url{ https://arxiv.org/abs/1801.07661}

je trouve pas ça super original mais je suis peut-être médisant}

\feedbackNIOK{Jerome Durand-Lose}{
- références à retravailler (champs manquant, double…)
H.}

\feedbackNIOK{JDL}{Voir document annoté à la main: intrégré plusieurs
  trucs. Pas tout car pas tout compris. }
\feedbackOK{Klaus Meer}{
I personnaly would use capital letters for complexity classes and
similar things. Example PCP instead of pcp, DNA instead of Dna, etc

I also think you should update many referencs. 
}
\olivierOK{IL FAUT CONTINUER A NETTOYER LA BIBLIO}

\coFINAL{\newpage \tableofcontents}
\FINAL{\bibliographystyle{spmpsci_cca}}
\coFINAL{\bibliographystyle{apalike}}

\bibliography{cca,Bib_BP,Bib_BP2} 

\end{document}